\documentclass[journal]{IEEEtran}
\usepackage{lipsum}
\usepackage{stfloats}
\usepackage{cite}
\usepackage{amsmath}
\usepackage{amsfonts}
\usepackage{balance}
\usepackage{fancyhdr}
\usepackage{graphicx,times,cite,amssymb,epsfig,amsthm,color}
\usepackage[normalem]{ulem}
\usepackage{xcolor}
\usepackage{diagbox}
\usepackage{makecell}   

\theoremstyle{remark}
\newtheorem{remark}{Remark}

\theoremstyle{remark}
\newtheorem{result}{Result}

\newlength{\figwidth}
\renewcommand{\baselinestretch}{1}    
\normalsize

\usepackage{graphicx}
\ifCLASSINFOpdf
\else
\fi
\usepackage{amsmath}
\DeclareMathOperator*{\argmax}{argmax}

\usepackage[cmintegrals]{newtxmath}
\usepackage{algorithm}
\usepackage{algorithmic}

\usepackage{array}

\usepackage{stfloats}
\usepackage{url}

\usepackage{verbatim}

\newcommand{\numfreqs}{M}
\newcommand{\numslots}{L}

\newcommand{\pri}{T}

\newcommand{\bw}{B}
\newcommand{\freqsep}{\Delta f}

\newcommand{\pulset}{p}

\newcommand{\uspec}{U}
\newcommand{\eqsymbol}[1]{\mathrel{\overset{\makebox[0pt]{\mbox{\normalfont\tiny\sffamily #1}}}{=}}}

\begin{document}
%
\title{Dynamic Length FSK Waveforms for Joint Communications and Radar}

%
%
%

\author{Tian~Han,~\IEEEmembership{Member,~IEEE,}
        Peter~J~Smith,~\IEEEmembership{Fellow,~IEEE,}
        Urbashi~Mitra,~\IEEEmembership{Fellow,~IEEE,}
        Jamie~S~Evans,~\IEEEmembership{Senior~Member,~IEEE,}
        Robin~Evans,~\IEEEmembership{Life~Fellow,~IEEE,}
        and~Rajitha~Senanayake,~\IEEEmembership{Member,~IEEE}
        

\thanks{Tian Han, Jamie S. Evans, Robin Evans and Rajitha Senanayake are with the Department of Electrical and Electronic Engineering, University of Melbourne, VIC 3010, Australia (e-mail: \{tian.han1, jse, robinje, rajitha.senanayake\}@unimelb.edu.au).}

\thanks{Peter J. Smith is with the School of Mathematics and Statistics, Victoria University of Wellington, Wellington, 6012, New Zealand (e-mail: peter.smith@msor.vuw.ac.nz).}

\thanks{Urbashi Mitra is with the Department of Electrical and Computer Engineering, University of Southern California, CA 90089, USA (e-mail: ubli@usc.edu).}
}

\maketitle

\begin{abstract}
Motivated by the constant modulus property of frequency shift keying (FSK) based waveforms and the stabilisation of its radar performance with an increase in the number of subpulses, in this paper an FSK-based dynamic subpulse number joint communications and radar waveform design is proposed. From a communications point of view, the system operates based on traditional FSK modulation. From a sensing point of view, although the subpulses are continuously generated and transmitted, radar waveforms are dynamically formed by monitoring the flatness of the spectrum which in return guarantees the accuracy of the delay estimation. Other constraints on the waveform length are used to ensure satisfactory values of the root mean square time duration, ambiguity function sidelobe levels and prevent overly long waveforms. To provide an estimation of the probability of generating extremely long waveforms, the distribution of the number of subpulses is approximated using a Brownian motion process and an existing result on its one-sided exit density. Numerical examples are provided to evaluate the accuracy of the approximate distribution, as well as the ambiguity function sidelobe levels and the delay and Doppler shift estimation performance of the transmitted waveforms.
\end{abstract}


%
\IEEEpeerreviewmaketitle

\section{Introduction} \label{sec:ch_adpt_mtvt_ctbt}
To support potential future application scenarios in which both sensing and communications play critical roles, integrated sensing and communications (ISAC) and its mainstream, joint communications and radar (JCR), are considered key enablers in next-generation wireless networks \cite{liu22abb, Wei23, kaushik2023isac}. To achieve such integration, a critical challenge is the design of a unified waveform that can perform sensing and data transmission simultaneously. Motivated by the unit baseband peak-to-average power ratio (PAPR) property of FSK and its radar performance analysed in \cite{Han23local,Han25global}, in this paper we explore the FSK-based JCR waveform design. Although FSK is a traditional communication modulation scheme, the idea of applying it to JCR originated from traditional radar stepped frequency waveforms \cite{Levanon2004}. Hence, we consider FSK-based JCR as a radar-centric design method. Due to its random nature, using fixed-length FSK-based waveforms leads to uncontrolled radar sensing performance \cite{Han23local,Han25global}. To address this issue, in this paper we consider using dynamic numbers of subpulses in order to guarantee particular radar performance metrics. The use of variable subpulse numbers is motivated by adaptive radars, which usually design waveforms or parameters based on prior knowledge of the radar sensing environment. In the following section, we first conduct a literature review on JCR waveform designs and adaptive radars.

\subsection{Literature Review}  \label{sec:literature}
JCR waveform designs are usually classified as communications-centric methods, radar-centric methods and designs from the ground up.  

Communications-centric methods incorporate the radar sensing functionality into existing communications waveforms. One attractive communications-centric JCR candidate is the orthogonal frequency division multiplexing (OFDM) waveform, which is widely used in cellular systems. In \cite{Gaudio19}, the maximum likelihood (ML) joint delay-Doppler estimator for OFDM is derived. Both the analytical Cramer-Rao lower bounds (CRLBs) and numerical results show that OFDM can achieve accurate radar estimation without affecting its high data rate. In \cite{Wang22}, the triangular frequency modulation is combined with OFDM to improve the delay Doppler estimation capabilities further on top of standard OFDM. 
While designs based on OFDM achieve a considerably high data rate, their high PAPRs are problematic for both radar sensing and communications as they introduce non-linear distortion at radio frequency front ends \cite{Huang22}. While extensive research has been conducted to reduce the PAPR of OFDM for pure communications systems \cite{Jiang08}, recently the design of low PAPR OFDM-based JCR has also attracted interest. In \cite{Huang22} and \cite{Piyush23}, a portion of the OFDM subcarriers is not used for data transmission but for minimising the PAPR instead. Such techniques can lead to low PAPR values if the number of subcarriers for PAPR reduction is large. However, this reduces the data rate significantly and requires the solution of non-trivial optimisation problems in real time.

While the PAPR issue has been raised in recent years for communications-centric designs, their radar-centric counterparts have focused on this from a very early stage. One main reason is that traditional radar systems usually use waveforms with a unit baseband PAPR, such as stepped frequency or phase coded ones \cite{richards14,Levanon2004}, to maximise the power efficiency of the power amplifier at the transmitter. Motivated by the ideal PAPR property and the deterministic time-bandwidth product control provided by using frequency steps, \cite{Senanayake22TWCabb} embeds data through frequency step permutations, achieving data transmission and maintaining good delay and Doppler shift estimation capabilities on average. To reduce the communications error rate or the radar ambiguity function (AF) peak sidelobe level (PSL), \cite{Dayarathna23} considers choosing subsets of frequency permutations carefully at a cost of data rate reduction. Incorporating phase shift keying (PSK) into stepped frequency-based designs improves their non-ideal data rate with little impact on the radar sensing capability \cite{Han2023_freqpermPSK}. Alternatively, \cite{Han25global} uses subpulse-wise independent FSK to increase the data rate while controlling the AF PSL by optimising the phase sequence. 

Waveform designs from the ground up do not rely on existing communications or radar waveforms. They are usually achieved by solving optimisation problems on sensing and communications performance metrics, allowing more flexibility in performance tradeoffs. Recently, the requirement for low PAPR has also attracted research interests under this framework. Such requirements are usually formulated as equality or inequality constraints, depending on whether a strictly unit PAPR is required. For example, \cite{Shi24} proposes a constant modulus waveform design for a reconfigurable intelligent surface (RIS)-assisted downlink ISAC system. The transmitted waveform and the RIS phase shifts are jointly designed to minimise the cross-correlation pattern among target signals, with inequality constraints on per-target illumination power and communications multi-user interference (MUI), and equality constraints for the unit PAPR requirement. In \cite{Bazzi23}, a downlink ISAC system with clutter is considered. The communications MUI is minimised to maximise the sum rate, with the PAPR and the similarity between the transmitted waveform and an ideal radar waveform bounded using inequality constraints. It is worth noting that these optimisation problems are usually non-convex, and their solutions have high complexity and/or require reformulation.

The use of dynamic waveform parameters in this work is initially motivated by adaptive radar systems, in which the succeeding waveforms are designed to enhance the sensing performance based on prior knowledge of the environment. One specific type of adaptive radar changes the parameters of the next waveform, including pulse repetition interval, coherent processing interval, pulse width, number of pulses, pulse shaping function and carrier frequencies of pulses, using previous estimates to achieve a set of system requirements \cite{greco18}. The system requirements can usually be interpreted as optimisation problems, aiming at minimising estimation and tracking errors \cite{Ghadian2020adaptive} and/or reducing interference \cite{Steck18}. Other types of adaptive radars include selecting a class of waveforms to decrease the estimation mean squared error (MSE) \cite{Tsistrakis2014} and designing the waveform specifically by solving optimisation problems to suppress the clutter \cite{Ying2009}, increase the detection probability \cite{Zhang2018_adaptive} or increase mutual information (MI) between the target and the observations \cite{Haykin2010}. The latter has some similarity to resource allocation problems in JCR waveform designs, as both rely on solving optimisation problems \cite{prelcic2024survey}.
Nevertheless, adaptive radar waveforms are designed solely based on the radar sensing environment and cannot transmit data; while optimisation-based JCR waveform designs usually consider optimising both radar and communications, or optimising one while constraining the other. 

\subsection{Motivation and Contributions} \label{sec:contribution}
In this work, we propose an FSK-based dynamic waveform length scheme for simultaneous communications and radar sensing. The use of FSK for JCR was initially motivated by the radar sensing capability and constant modulus property of stepped frequency radar waveforms \cite{Han23local, Han25global}. Conventional radar systems usually consider only constant modulus waveforms, such as frequency or phase coded ones \cite{Levanon2004}. This is because power amplifiers in radar transmitters usually operate at or near the saturation point to maximise the transmit power and thus, improve the detection and estimation capabilities. When a waveform with a PAPR greater than one passes through, it suffers from non-linear distortion, which degrades its practical radar performance \cite{palo2020}. In particular resource-limited ISAC scenarios, such as internet of things (IoT) applications, where power or cost efficiency is prioritised over spectral efficiency, the low or even unit baseband PAPR property can be essential \cite{akan2020}. Examples include low-power wireless sensor networks \cite{Ali24}, particularly low-power wide-area networks \cite{Wu23lpwan}, where remote sensing should be achieved using small devices powered by batteries with limited capacity. Such applications prefer power and cost efficiency over high spectral efficiency, making low or unit PAPR schemes such as FSK suitable candidates, as they efficiently make use of the limited transmit power. This is evident from an LPWAN technology called Long Range (LoRa), which uses a modified version of FSK as the physical layer modulation \cite{de20195g}. On the other hand, the high PAPR value of OFDM-typed waveforms leads to out-of-band and in-band distortions when passing through non-linear devices in transceivers \cite{Jiang08}. The former induces leakage to adjacent channels, while the latter leads to a drop in the in-band signal-to-noise ratio (SNR). PAPR reduction techniques, such as selective mapping and digital predistortion, can be used to lessen these problems, but they introduce additional complexity burdens and/or performance degradations. Although OFDM is suitable and has been used extensively in cellular systems, either the high PAPR or the additional complexity required for PAPR reduction makes OFDM unsuitable for the aforementioned resource-limited scenarios.

Though the unit baseband PAPR property makes FSK suitable for resource-limited ISAC scenarios, it suffers from having uncontrolled radar sensing capability due to its random nature \cite{Han23local, Han25global}. To address this issue, in this work we propose waveforms with dynamic numbers of subpulses to guarantee specific radar performance measures. The use of dynamic waveform parameters has similarities to adaptive radar systems. Nevertheless, adaptive radar waveforms designed solely based on the radar-sensing environment cannot transmit data; while we consider a dynamic parameter system in which only the number of radar waveform subpulses is dynamically changed. More specifically, the transmitter continuously generates and transmits single carrier subpulses based on $\numfreqs$-ary FSK. The communications receiver performs symbol-by-symbol (i.e., subpulse-by-subpulse) FSK detection. At the radar receiver, the waveform processed is formed by dynamically selecting a contiguous sequence of those continuously generated subpulses by monitoring certain radar performance metrics until pre-defined requirements are satisfied. Such a scheme has the same communications modulation and demodulation operations and performance as a typical FSK scheme from the communications perspective while guaranteeing the radar performance of interest for each processed waveform from the radar perspective.

The dynamic length of the resulting radar waveform can be beneficial or disadvantageous to the radar performance. In typical radar applications \cite{Paulose1994, Soares1996, Kozlov22}, waveforms usually have several hundreds to thousands of subpulses. The use of such long waveforms is advantageous in terms of the Doppler estimation accuracy which is inversely related to the time duration occupied by a waveform \cite{Han23local}. For the traditional stepped frequency waveform or the linear frequency modulated (LFM) waveform \cite{Levanon2004}, it also guarantees a large bandwidth, leading to good delay estimation performance. Moreover, the ambiguity function (AF) sidelobe levels (SL) of particularly designed frequency sequences, such as Costas codes, are less than or equal to the reciprocal of the number of subpulses \cite{Levanon2004}, indicating the advantage of using long waveforms in terms of clutter mitigation. 

On the other hand, excessively long waveforms can lead to significant delays in radar sensing, posing a practical challenge. Hence, it is reasonable to set an upper limit on the length, so that certain desirable radar properties are maintained but the waveform is not too long. In this work, we derive the distribution of the subpulse number so that we can assess the probability that the waveform length lies in the desired region. More specifically, we focus on the flatness of the frequency spectrum of each processed radar waveform to evaluate the subpulse number distribution, as the other proposed metrics to control the radar performance give deterministic bounds as shown in Section \ref{sec:perf_req}. The use of spectrum flatness is motivated by traditional frequency-modulated radar waveforms such as stepped frequency ones, whose frequency spectra are uniform across the available bandwidth, leading to preferable radar performance, especially the delay estimation capability \cite{Levanon2004}. 
Our main contributions are summarised in the following.
\begin{itemize}
    \item We propose a novel and simple dynamic length FSK-based JCR scheme, which guarantees the radar performance while not affecting the communications functionality. This is achieved by monitoring radar performance metrics of interest until pre-defined requirements are satisfied when forming the radar waveform.

    \item We demonstrate that commonly used radar performance metrics of the FSK-based waveform, including AF SLs and the delay-Doppler estimation capabilities \cite{Levanon2004} either stabilise to values that can be controlled by tuning waveform parameters or improve with an increase in the number of subspulses. 

    \item We analyse the probability distribution of the subpulse number for the scheme in which the frequency spectrum flatness is the metric to be monitored. Since the metric is monitored while generating subpulses, we translate the problem to finding the distribution of the hitting time of a stochastic spectrum flatness process and derive an accurate approximation to the distribution.

    \item We provide numerical examples to analyse the accuracy of the proposed approximation. In addition, the performance of the dynamic subpulse number scheme is assessed in terms of the AF SL and the delay and Doppler estimation mean squared errors (MSEs).
\end{itemize}
The rest of the paper is organised as follows. Section \ref{sec:ch_adpt_sig_mod} proposes the system and signal models, and discusses the radar performance metrics used in the paper. Section \ref{sec:sp_no} analyses the probability distribution of the number of subpulses when the frequency spectrum flatness is the metric to be monitored. Section \ref{sec:ch_adpt_num_examples} presents numerical examples to support the analysis and compares the radar estimation performance of the proposed dynamic scheme with its fixed-length counterpart. Finally, Section \ref{sec:conclusion} provides conclusions and potential future extensions of this work.

\section{Problem Formulation} \label{sec:ch_adpt_sig_mod}

\subsection{System Model}
\begin{figure}[h]
    \centerline{\includegraphics[width=8.5cm]{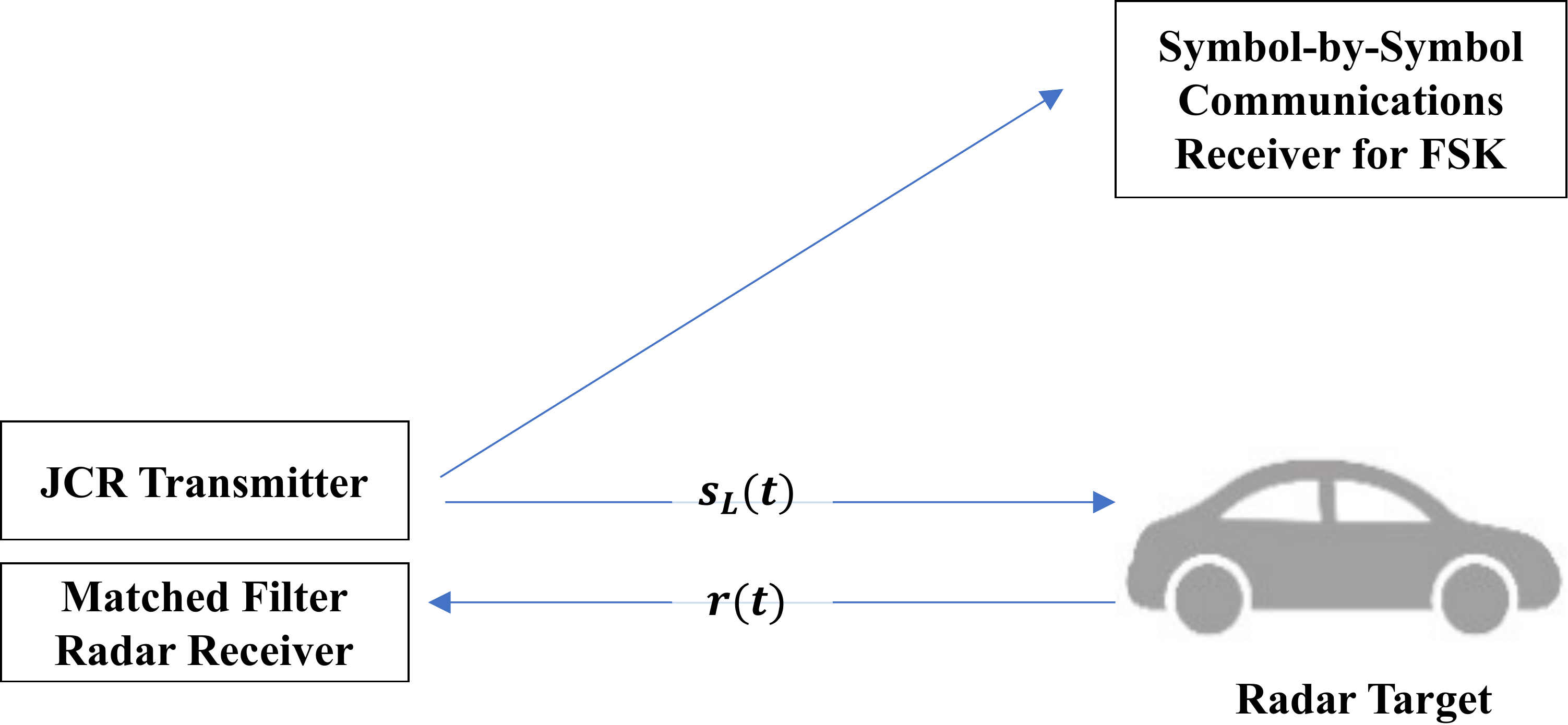}}
    \caption{The JCR system model.} \label{fig:sys_model}
\end{figure}

Fig. \ref{fig:sys_model} describes the JCR system model used in this work. At the JCR transmitter, a waveform consisting of multiple FSK-modulated subpulses is sent for both radar target sensing and data transmission. The signal received at the communications receiver is processed using the symbol-by-symbol ML detector to obtain the transmitted data \cite{Proakis08}. The signal bounced back by the radar target is received by the radar receiver co-located with the transmitter and processed to estimate the relative range and velocity of the target. Assume that the complex envelope of the radar received signal reflected by a single target can be expressed as \cite[eq.(10.5)]{vantrees01}
\begin{align}
    r(t) = \Tilde{b} s(t-\tau) e^{j\omega t} + n(t),
\end{align}
where $s(t)$ is the complex envelope of the transmitted signal, $\tau$ represents the round trip delay related to the relative range, $\omega$ is the Doppler frequency shift caused by the relative velocity, $\Tilde{b}$ is a complex Gaussian random variable with zero mean and variance $\sigma^2_{\Tilde{b}}$ representing the reflection from multiple reflecting surfaces on the target and $n(t)$ is the baseband complex additive white Gaussian noise (AWGN) process with zero mean and power spectral density (PSD) $N_0$. The delay and Doppler shift can be estimated using the ML matched filter estimator as \cite[eq.(10.6)]{vantrees01}
\begin{equation}
    \left(\hat{\tau},\hat{\omega}\right) = \argmax_{(\Bar{\tau},\Bar{\omega})} \left\{ \left|\int_{-\infty}^{\infty} r(t) s^*(t-\Bar{\tau})e^{-j\Bar{\omega} t} d t \right|^2\right\}, \label{eq:cont_est}
\end{equation}
where $\hat{\tau}$ and $\hat{\omega}$ are estimates of the delay and Doppler shift, respectively. In practice, a discrete-time approximation of \eqref{eq:cont_est} is usually considered. 

\subsection{Signal Model}
Consider a JCR transmitter which continuously transmits single-tone subpulses whose centre frequency is decided based on the incoming data stream using $\numfreqs$-ary FSK modulation. The communications receiver processes the received symbols to detect the embedded data bits as in traditional FSK communications systems. From the radar perspective, however, when each subpulse is generated, some radar performance metrics of the waveform composed of all generated subpulses after the previously processed radar waveform 
are evaluated. Once all those metrics satisfy pre-defined requirements, the entire waveform is regarded as the radar sensing waveform and processed at the radar receiver for target delay and Doppler estimation. Hence, from the radar perspective, the number of subpulses, $\numslots$, is not fixed but dynamic, determined based on pre-defined performance requirements. 
The complex envelope of the JCR waveform can be expressed as
\begin{align}  \label{eq:sig_model}
    s_{\numslots}(t) = \sum_{l=0}^{\numslots-1} p (t-l\pri) \exp(j \omega_l(t-l\pri)),
\end{align}
where $\numslots$ is the number of the subpulses, $\pri$ is the subpulse repetition interval (SPRI) and $p(t)$ is a rectangular pulse which can be expressed as
\begin{align} \label{eq:rect_pulse}
\pulset(t) = \left\{
\begin{array}{ll}1~~~~~~~~&0 \leq t \leq \pri \\
0~~~~~~~~&\text{otherwise.} 
\end{array}\right.
\end{align}
The centre frequency of the $l$-th subpulse, $\omega_l$, is decided based on $\numfreqs$-ary FSK, i.e., $\omega_l \in \{2\pi f_0, \cdots, 2\pi (f_0+(\numfreqs-1)\Delta f)\}$, where $f_0$ is the starting frequency and $\Delta f = i/T$ with non-zero integer $i$ is the separation between adjacent frequencies. We assume that $\Pr[\omega_l = 2\pi (f_0 + m \Delta f)] = 1/\numfreqs, \forall m \in \{0, \cdots, \numfreqs-1\}$ and $\omega_l, l \in \{0,\cdots, \numslots-1\}$ are mutually independent. To emphasise the dynamic subpulse number, a subscript $\numslots$ is introduced to the waveform notation $s_{\numslots}(t)$. The number of subpulses is decided as the first $\numslots$ that satisfies a set of inequalities, each representing a performance requirement, i.e., 
\begin{align}  \label{eq:ch_adpt_pfmc_rqmts}
    \numslots = \min \left\{\Tilde{\numslots} \in \mathbb{Z}^+ \,\,\vline\,\, g_j(\Tilde{L}) \geq \gamma_j, \forall j \in \{1,\cdots, J\} \right\}
\end{align}
where $g_j(\Tilde{\numslots})$ is the $j$-th performance metric which depends on the sequence of $\Tilde{\numslots}$ subpulses, $\gamma_j$ is the threshold associated with the requirement of the $j$-th metric, and $J$ is the total number of metrics we consider. The performance metrics will be related to the radar performance required. Also, they should be simple enough so monitoring them after each subpulse is computationally inexpensive. Detailed discussions of potential performance metrics and requirements are provided later in Section \ref{sec:perf_req}.

For particular performance requirements, the dynamic length FSK scheme described in \eqref{eq:sig_model}-\eqref{eq:ch_adpt_pfmc_rqmts} can lead to the use of long waveforms. In typical implementations of radar only applications, a waveform usually has several hundreds to thousands of subpulses, depending on the requirements of particular applications \cite{Paulose1994, Soares1996, Kozlov22}. As has been discussed in Section \ref{sec:contribution}, when frequency modulated waveforms particularly designed for radar applications \cite{Levanon2004} are considered, using such long sequences enhances both the delay Doppler resolutions and the AF SLs. In JCR systems, however, the requirement of data transmission introduces a random nature. Nevertheless, when FSK-based JCR schemes are considered, using a large number of subpulses is still beneficial for the aforementioned radar performance metrics. 
These conclusions are drawn from the comprehensive discussion and analysis provided in Appendix \ref{sec:ch_adpt_large_no_subpulses}, and are listed in the following:
\begin{itemize}
\item Appendix \ref{sec:clutter_afsl} indicates that the AF SL at a particular position on the delay-Doppler plain depends on the sequence of embedded FSK symbols. With an increasing subpulse number $\numslots$, it stabilises to its statistical mean, whose value does not depend on $\numslots$. 
\item Appendix \ref{sec:dop_time_del_bw} indicates that the Doppler shift estimation capability, represented by the root mean square (RMS) time duration of the waveform, is independent of the frequency sequence. It improves monotonically with $\numslots$.
\item Appendix \ref{sec:dop_time_del_bw} also indicates that the delay estimation capability, represented by the RMS bandwidth of the waveform depends on the frequency sequence. With increasing $\numslots$, it stabalises towards a situation where the frequencies in the sequence are almost evenly distributed among the available bandwidth.
\end{itemize}

\begin{figure}[t]
    \centerline{\includegraphics[width=9cm]{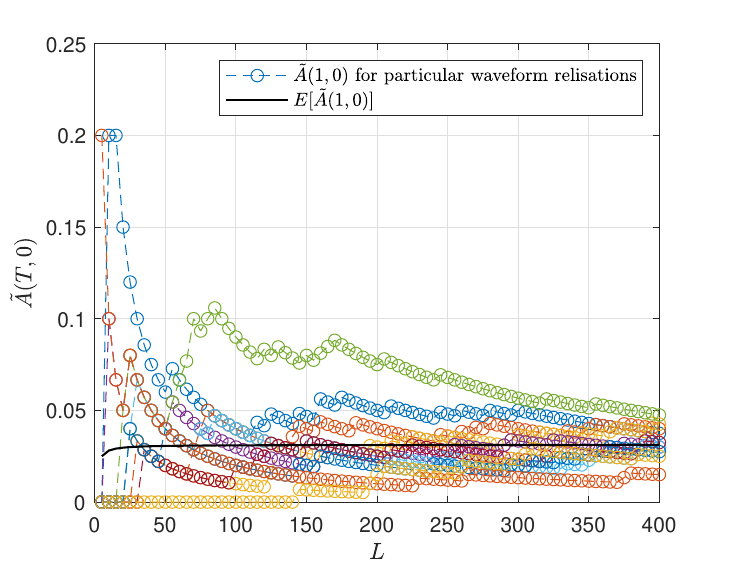}}
    \caption{The AF SL at the delay-Doppler pair $(\tau,\omega) = (1 \times T, 0)$ versus $\numslots$, for $\numfreqs = 32$.} \label{fig:AFSL_vs_L}
\end{figure}

Here, we also provide one numerical example to support the statement. Fig. \ref{fig:AFSL_vs_L} plots the AF SL at the delay-Doppler pair $(\tau,\omega) = (1 \times T, 0)$, denoted by $\Tilde{A}(1,0)$, versus the subpulse number $\numslots$. The AF SLs for multiple frequency sequence realisations are plotted as coloured dashed lines with circle markers. The average value of $\Tilde{A}(1,0)$, i.e., $E[\Tilde{A}(1,0)]$, whose expression is given in \eqref{eq:AF_kr_mean} in Appendix \ref{sec:clutter_afsl}, is plotted as a black solid line as a reference. We can clearly observe that the variance of $\Tilde{A}(1,0)$ decreases with an increase in $\numslots$, indicating that for large $\numslots$, the AF SL stabilises to its mean value. 

\subsection{Performance Requirements for the Dynamic Scheme} \label{sec:perf_req}
The analysis in Appendix \ref{sec:ch_adpt_large_no_subpulses} clearly shows that typical radar performance metrics either improve to or stabilise at an acceptable level with $\numslots$. Nevertheless, this performance stabilisation is in a probabilistic manner, indicating that fixed length FSK schemes cannot guarantee the performance of each single radar waveform. The proposed novel dynamic length scheme in \eqref{eq:sig_model} to \eqref{eq:ch_adpt_pfmc_rqmts}, on the other hand, can adopt the benefits of large subpulse numbers while guaranteeing certain radar performance for each single radar waveform. Now we need to decide what important performance requirements should be considered in \eqref{eq:ch_adpt_pfmc_rqmts}. Based on Appendix \ref{sec:ch_adpt_large_no_subpulses}, the performance metrics we can consider to determine the value of $\numslots$ are discussed in the following. 

\subsubsection{RMS Time Duration} 
It is obvious from \eqref{eq:secdop_J22} in Appendix \ref{sec:dop_time_del_bw} that the RMS time duration increases quadratically with the number of subpulses. Therefore, a specific requirement on the RMS time duration, i.e., 
\begin{align} \label{eq:ch_adpt_rmstd_req_}
    \sigma_{t}^2 = \frac{\numslots^2\pri^2}{12} \geq \left(\sigma_{t}^2\right)_{\text{req}},
\end{align}
can be simply interpreted as a lower bound on $\numslots$, i.e, 
\begin{align}  \label{eq:ch_adpt_rmstd_req}
    \numslots \geq \gamma_1 = \sqrt{\frac{12\left(\sigma_{t}^2\right)_{\text{req}}}{\pri^2}}.
\end{align}

\subsubsection{Frequency Spectrum} 
Unlike the RMS time duration, an increase in $\numslots$ does not guarantee an increase in the RMS bandwidth. Nevertheless, the analysis in Appendix \ref{sec:dop_time_del_bw} shows that using a large enough $\numslots$ value leads to stabilisation towards a situation in which the centre frequencies are almost evenly distributed among the available spectrum. Using the spectrum flatness indicator defined in \eqref{eq:spctrm_fltns_mtrc} in Appendix \ref{sec:metric_spec_flat}, a performance requirement on it can be expressed as
\begin{align} \label{eq:ch_adpt_rmsbw_req}
    \uspec(\numslots) = \frac{\sum_{m=0}^{\numfreqs-1}\left(\frac{N_m(\numslots)}{L}-\frac{1}{\numfreqs}\right)^2}{\numfreqs} \leq \gamma_2,
\end{align}
where $\uspec(\numslots)$ is the MSE measuring the deviation of the proportions of the frequencies from the desired flat spectrum and $N_m(\numslots)$ denotes the number of the times the $m$-th frequency was selected in the sequence after $\numslots$ subpulses are generated. Slightly different from \eqref{eq:spctrm_fltns_mtrc}, in \eqref{eq:ch_adpt_rmsbw_req} we write both $\uspec$ and $N_m$ as functions of $\numslots$ to emphasise that their values are updated once a subpulse is generated. 

\subsubsection{AF Sidelobe Levels} 
Ideally, to monitor the capability of a waveform to mitigate clutter presenting at any delay-Doppler value, computing all AF SLs of the entire waveform after each subpulse is generated is required. However, this requires a large number of operations, as both the number of operations to compute each AF SL and the number of SLs to be monitored increase with $L$. More specifically, $\Tilde{A}(k,r)$ is a summation of $L-k$ addends, and the cardinality of the set of $(k,r)$ values is $|\mathcal{D}| = 2\numslots\numfreqs+\numslots+\numfreqs$, as shown in Appendix \ref{sec:clutter_afsl}. Due to the high computational complexity, performing online tracking of AF SLs after each subpulse is generated is extremely difficult. Hence, instead we go for a feasible and simple approach. As shown by both Fig. \ref{fig:AFSL_vs_L} and Appendix \ref{sec:clutter_afsl}, the AF SLs stabilise to their mean values when the waveform gets longer, since their variances decrease with increasing $L$. Hence, to control the variances of all SLs, we consider the largest variance over different $(k,r) \in \mathcal{D}$ pairs. It is clear from \eqref{eq:AF_kr_var} that for any particular $\numslots>0$ and $\numfreqs \geq 2$, $\text{Var}\left[\Tilde{A}(k,r;\numslots)\right]$ decreases with both $k$ and $|r|$. Note that we introduce an extra argument $\numslots$ to the AF SL $\Tilde{A}(\cdot)$ since $\numslots$ is not fixed. Therefore, we determine $\numslots$ based on the largest variance $\text{Var}\left[\Tilde{A}(0,1;\numslots)\right]$. Generally speaking, if the number of subpulses, $\numslots$, is selected such that the variance $\text{Var}\left[\Tilde{A}(0,1;\numslots)\right]$ satisfies
\begin{align} \label{eq:ch_adpt_slvar_req_}
    \text{Var}\left[\Tilde{A}(0,1;\numslots)\right] \leq \left(\sigma_{A}^2\right)_{\text{req}},
\end{align}
then the variances of all other SLs are smaller than $\left(\sigma_{A}^2\right)_{\text{req}}$. Using \eqref{eq:AF_kr_var}, the requirement in \eqref{eq:ch_adpt_slvar_req_} can be interpreted as
\begin{align} \label{eq:ch_adpt_slvar_req}
    \numslots \geq \gamma_3 = \frac{(\numfreqs-1)(\numfreqs^2-\numfreqs+1)}{\numfreqs^4 \left(\sigma_{A}^2\right)_{\text{req}}}.
\end{align}

\subsubsection{Processing Latency} 
Apart from the requirements on the aforementioned performance measures, in practice we also need to monitor the length of each waveform. Although long waveforms have various beneficial radar properties, arbitrarily long waveforms are not acceptable as these introduce excessive processing latency to radar sensing. Therefore, we consider a simple upper bound on the number of subpulses, i.e.,
\begin{align} \label{eq:ch_adpt_length_req}
    \numslots \leq \gamma_4. 
\end{align}

\subsubsection{Complexity Discussion and Conclusions}
Although there are requirements on the RMS time duration, the AF SLs and the time required to form a waveform, \eqref{eq:ch_adpt_rmstd_req}, \eqref{eq:ch_adpt_slvar_req} and \eqref{eq:ch_adpt_length_req} indicate that these requirements can be interpreted as simple upper and lower bounds on $\numslots$. Thus, they do not need to be monitored throughout the process in real time. Unlike the others, the spectrum flatness requirement in \eqref{eq:ch_adpt_rmsbw_req} does not give a simple inequality for $L$. Fortunately, the computation of $\uspec(\numslots)$ is easy. We assume that the values of $\numslots$ and $N_m(\numslots), \forall m \in \{0,\cdots,\numfreqs\}$ are stored and updated once a subpulse is generated. The update includes adding $1$ to both $\numslots$ and the $N_m(L)$ corresponding to the latest frequency symbol. Rewriteing $\uspec(L)$ in \eqref{eq:ch_adpt_rmsbw_req} as
\begin{align} \label{eq:UL}
    \uspec(\numslots) = \frac{\sum_{m=0}^{\numfreqs-1}\left(N_m(\numslots)-\frac{\numslots}{\numfreqs}\right)^2}{\numslots^2\numfreqs},
\end{align}
it is straightforward to show that computing $U(L)$ requires $3M-1$ additions and $3M+3$ multiplications. Rearranging the expression in \eqref{eq:UL} can lead to higher or lower numbers of arithmetic operations, but generally speaking, the numbers are $\mathcal{O}(M)$.

Since excessively long waveforms can lead to significant periods of time without radar sensing, an upper limit of $\numslots \leq \gamma_4$ is proposed in \eqref{eq:ch_adpt_length_req}. Also, control of AF SLs and RMS time duration gives $\numslots \geq \max(\gamma_1 , \gamma_3)$. As these three metrics all give simple deterministic bounds on $\numslots$, we focus on analysing the random subpulse number required to guarantee a flat spectrum. More specifically, we analyse the probability distribution of the first $\numslots$ that satisfies \eqref{eq:ch_adpt_rmsbw_req}. We note that if the first $L$ value that meets \eqref{eq:ch_adpt_rmsbw_req} also satisfies $\max(\gamma_1 , \gamma_3) \leq \numslots  \leq \gamma_4$, all the radar constraints are simultaneously satisfied.

\section{Subpulse Number Distribution Analysis} \label{sec:sp_no}
In this section, we analyse the number of subpulses required to guarantee a flat spectrum. If we only monitor the criterion in \eqref{eq:ch_adpt_rmsbw_req}, the waveform is ready to be processed by the radar receiver once the spectrum flatness requirement is satisfied. More specifically, the number of subpulses should be the first $\numslots$ that satisfies \eqref{eq:ch_adpt_rmsbw_req}. In other words, it is the time for the stochastic process, $\uspec(\numslots)$, to hit a pre-defined threshold, $\gamma_2$. We note from \eqref{eq:ch_adpt_rmsbw_req} that $\uspec(\numslots)$ has some similarity with the well-known chi-squared test statistic, which is a measure of the difference between the observed and expected frequencies of outcomes of a set of events or variables. More specifically, noting that the expected number of each centre frequency is $\numslots/\numfreqs$, the chi-squared test statistic for $N_m(\numslots), m \in \{0, \cdots, \numfreqs-1\}$ can be expressed as \cite{Lawal1980}
\begin{align}   \label{eq:ch_adpt_chisq}
    \chi^2(\numslots) = \frac{\sum_{m=0}^{\numfreqs-1}\left(N_m(\numslots)-\frac{\numslots}{\numfreqs}\right)^2}{\frac{\numslots}{\numfreqs}}.
\end{align}
Note that $\chi^2(\numslots) = \numslots \numfreqs^2 \uspec(\numslots)$. The chi-squared test statistic, $\chi^2(\numslots)$, is asymptotically chi-squared distributed with $M-1$ degrees of freedom for large $\numslots$. If $\numfreqs$ is also large, $\chi^2(\numslots)$ is a sum of a large number of independent random variables, and thus is approximately Gaussian distributed based on the CLT, i.e.
\begin{align} \label{eq:ch_adpt_chi2_Gau}
    \frac{\chi^2(\numslots)-E\left[\chi^2(\numslots)\right]}{\sqrt{\text{Var}\left[\chi^2(\numslots) \right]}} \rightarrow \mathcal{N}(0,1), \text{ as } \numslots, \numfreqs \rightarrow \infty.
\end{align}
As discussed in Section \ref{sec:ch_adpt_sig_mod}, the stabilisation to a large RMS bandwidth and low AF SLs with increasing $\numslots$ requires a large $\numfreqs$ value. Hence, using a large $\numfreqs$ is sensible, making the Gaussian approximation of $\chi^2(\numslots)$ more accurate. We will also show this using numerical examples in Section \ref{sec:ch_adpt_num_examples}.

In general, it is difficult to solve the hitting time distribution of a stochastic process, except for special processes such as Brownian motion \cite{Ferebee82}. Since $\chi^2(\numslots)$ is approximately a Gaussian process for large $\numslots$, it can be shown that a linkage to the Brownian motion process is possible. To obtain this relationship, the mean, the variance and the correlation between time samples of $\chi^2(\numslots)$ are required. These required statistics of $\chi^2(\numslots)$ are derived and provided in Appendix \ref{app:autostats_chi}. Based on the CLT in \eqref{eq:ch_adpt_chi2_Gau} and the statistics in \eqref{eq:ch_adpt_E_chisq}, \eqref{eq:ch_adpt_var_chisq} and \eqref{eq:ch_adpt_cor_chisq}, we approximate $\chi^2(\numslots)$ as in Remark \ref{remark:chi2_Gau}. Note that the autocorrelation function used in Remark \ref{remark:chi2_Gau} is an approximation for large $\numslots$ values, as its exact expression makes the linkage between $\chi^2(\numslots)$ and a Brownian motion process overly difficult. 

\begin{remark} \label{remark:chi2_Gau}
For large $\numslots$ and $\numfreqs$, the stochastic process $\chi^2(\numslots)$ is approximately Gaussian, i.e., $\chi^2(\numslots) \sim \mathcal{N}(\numfreqs-1, 2(\numfreqs-1)(1-1/\numslots))$, with an auto-correlation function $\rho[\chi^2(\numslots), \chi^2(\numslots+k)] \approx \numslots/(\numslots+k)$.
\end{remark}

The Gaussian approximation and the correlation in Remark \ref{remark:chi2_Gau} allow us to link $\chi^2(\numslots)$ with a Brownian motion process. Since Brownian motion processes are continuous-time, we consider a continuous-time approximation of $\chi^2(\numslots)$, i.e., 
\begin{align}
    &\chi^2(t) \sim \mathcal{N}\left(\numfreqs-1, 2(\numfreqs-1)\left(1-\frac{1}{t}\right)\right), \label{eq:ch_adpt_chisq_ctns_dist} \\
    &\rho[\chi^2(t), \chi^2(t+t')] \approx \frac{t}{t+t'},   \label{eq:ch_adpt_chisq_ctns_corr}
\end{align} 
for large $t$ and $\numfreqs$.

Using Remark \ref{remark:chi2_Gau} and \eqref{eq:ch_adpt_chisq_ctns_dist}-\eqref{eq:ch_adpt_chisq_ctns_corr}, we obtain the following main result.
\begin{result} \label{result:ch2_BM}
The stochastic process $\chi_{\text{BM}}^2(t)$ defined by
\begin{align} \label{eq:ch_adpt_chisq_BM}
    \chi_{\text{BM}}^2(t) =  \numfreqs - 1 + \sqrt{2(\numfreqs-1)\left(1-\frac{1}{t}\right)} \frac{W(t^2)}{t},
\end{align}
where $W(t)$ is a Brownian motion process characterised by 
\begin{align} 
    W(t) &\sim \mathcal{N}(0, t), \label{eq:ch_adpt_BM_dist} \\
    E[W(t_1)W(t_2)] &= \min\{t_1, t_2\}, \label{eq:ch_adpt_BM_corr}
\end{align}
has the same mean, variance, distribution and autocorrelation function as the continuous-time approximation, $\chi^2(t)$, defined in \eqref{eq:ch_adpt_chisq_ctns_dist}-\eqref{eq:ch_adpt_chisq_ctns_corr}.
\end{result}
The proof of Result \ref{result:ch2_BM} is provided in Appendix \ref{app:proof_result}.

Using Result \ref{result:ch2_BM}, we approximate $\chi^2(t)$ for large $t$ and $\numfreqs$ as
\begin{align}    \label{eq:ch_adpt_chisq_BM_link}
    \chi^2(t) \approx \chi_{\text{BM}}^2(t).
\end{align}
The accuracy of the model will be discussed further using numerical examples in Section \ref{sec:ch_adpt_num_examples}. 

With the approximation in \eqref{eq:ch_adpt_chisq_BM_link}, we are able to solve the problem in \eqref{eq:ch_adpt_rmsbw_req} using an existing hitting time solution for Brownian motion processes \cite{Ferebee82}. More specifically, the PDF of the time, $t_0$, that the Brownian motion, $W(t)$, first hits a time-varying boundary, $b(t)$, can be approximated by \cite{Ferebee82}
\begin{align}  \label{eq:ch_adpt_tgt_aprx}
    f_{t_0}(t) \approx \frac{1}{\sqrt{2\pi t}} \exp\left(-\frac{b^2(t)}{2t}\right) \left(\frac{b(t)}{t} - \frac{d b(t)}{d t}\right), \quad t > 0.
\end{align}
The approximation in \eqref{eq:ch_adpt_tgt_aprx} is called the tangent approximation to the one-sided Brownian exit density. Following the detailed steps in Appendix \ref{app:hittime_pdf}, the PDF of the hitting time that the process $\uspec(t)$ satisfies \eqref{eq:ch_adpt_rmsbw_req}, $t_0$, can be approximated as
\begin{align}   \label{eq:ch_adpt_t_tgt_aprx}
f_{t_0}(t) & \approx \frac{1}{\sqrt{2\pi}} \frac{\left(\frac{\gamma_2\numfreqs^2}{2}-(\numfreqs-1)\right)t+\frac{\numfreqs-1}{2}}{\sqrt{2(\numfreqs-1)}\left(1-\frac{1}{t}\right)^{3/2}t^2} \notag \\ 
&\times \exp\left(-\frac{(\gamma_2\numfreqs^2t-(\numfreqs-1))^2}{4(\numfreqs-1)\left(1-\frac{1}{t}\right)}\right), \quad t > 0.
\end{align} 
Hence, the CDF of $t_0$ can be expressed as
\begin{equation}   
    F_{t_0}(t) \approx \int_{0}^{t}f_{t_0}(t) dt, \quad t > 0.
\end{equation}
The CDF of the number of subpulses, $\numslots_0$, that first satisfies the condition in \eqref{eq:ch_adpt_rmsbw_req} can be expressed as 
\begin{equation} \label{eq:ch_adpt_CDF_L}
    F_{\numslots_0}(\numslots) \approx F_{t_0}(\numslots), \quad \numslots > 0.
\end{equation}

The derived approximation of the subpulse number distribution in \eqref{eq:ch_adpt_t_tgt_aprx}-\eqref{eq:ch_adpt_CDF_L} provides an estimate of the probability that waveforms with a particular number of subpulses are used from the radar perspective. We will show using numerical examples that the approximation is accurate, especially for large $\numslots$ values. Although waveforms with large numbers of subpulses have stable radar sensing performance, extremely long waveforms can introduce excessive delays to radar sensing. This is handled by an upper bound on the length. Hence, an accurate estimate of the waveform distribution is useful to evaluate the probability that the length required for satisfactory spectrum flatness lies within the deterministic upper and lower bounds. Meanwhile, the approximation can be used as one potential guideline to be considered when deciding the design parameter $\gamma_2$. Suppose we want the probability of having waveforms with more than $\numslots_1$ subpulses to be less than $\alpha$. In this situation, we can approximate the range of $\gamma_2$ that satisfies this requirement by substituting \eqref{eq:ch_adpt_t_tgt_aprx}-\eqref{eq:ch_adpt_CDF_L} into $F_{\numslots_0}(\numslots_1) > 1-\alpha$ and solving this inequality numerically. 

\section{Numerical Examples} \label{sec:ch_adpt_num_examples}
In this section, we propose numerical examples to support our analysis of the hitting time and the performance of the processed waveform. 
Section \ref{sec:num_hit_time} focuses on validating the hitting time analysis, which only considers the spectrum flatness metric in \eqref{eq:ch_adpt_rmsbw_req}. 
Section \ref{sec:num_radar} presents the AF SLs (in Fig. \ref{fig:ch_adpt_A10_M32_gamma1e-4}) and delay-Doppler estimation capabilities (in Fig. \ref{fig:ch_adpt_MSE} and Fig. \ref{fig:Comp_SNR20}) of the proposed scheme, respectively. The impact of upper and lower bounds given by the requirements described in \eqref{eq:ch_adpt_rmstd_req}, \eqref{eq:ch_adpt_slvar_req} and \eqref{eq:ch_adpt_length_req} are included and discussed in Fig. \ref{fig:ch_adpt_MSE} and Fig. \ref{fig:Comp_SNR20}.

\subsection{Hitting Time Analysis Validation} \label{sec:num_hit_time}

\begin{figure}[h]
    \centerline{\includegraphics[width=9cm]{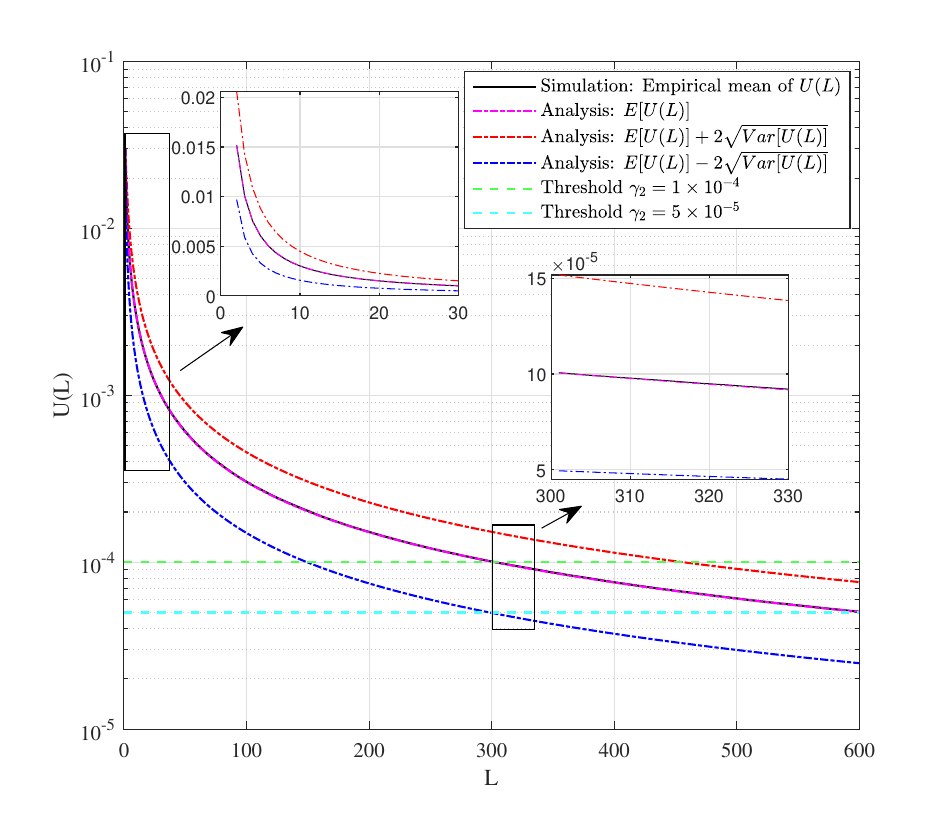}}
        \caption{The empirical mean, the analytical mean and the analytical mean $\pm$ two analytical standard deviations of $\uspec(\numslots)$ versus $\numslots$ for $\numfreqs=32$ and $\gamma_2 \in \{5\times 10^{-5},1\times10^{-4}\}$.} \label{fig:ch_adpt_mean_pm_std_M32_gamma1e-4}
\end{figure}

Fig. \ref{fig:ch_adpt_mean_pm_std_M32_gamma1e-4} plots the empirical mean, the analytical mean and the analytical mean $\pm$ two analytical standard deviations of the spectrum flatness measure, $\uspec(\numslots)$, versus $\numslots$ for $\numfreqs=32$ and $\gamma_2 \in \{5\times 10^{-5},1\times10^{-4}\}$. 
The analytical mean and standard deviation are calculated based on \eqref{eq:ch_adpt_E_chisq}, \eqref{eq:ch_adpt_var_chisq} and the relationship between $\chi^2(\numslots)$ and $\uspec(\numslots)$. We observe that the analytical mean accurately follows the empirical mean as expected. The analytical mean $\pm$ two standard deviation values are plotted to provide a range in which the majority of the $\uspec(\numslots)$ values are found. By zooming into the regions where $\numslots \in \{2,\cdots, 30\}$ and $\numslots \in \{301,\cdots, 330\}$, we observe that the standard deviation decreases rapidly with $\numslots$ in the low $\numslots$ region, while its value is extremely small in the high $\numslots$ region. This indicates that the spectrum flatness metric has only a small variation for a large $\numslots$ value. Note that the vertical axis of the main plot is logarithmic while the vertical axis of the zoom-in plots is on a linear scale. In addition, the intersections between the two standard deviation curves and the horizontal line representing the threshold provide us with a range in which the majority of the hitting time falls. For example when $\gamma_2 = 1 \times 10^{-4}$, we can conclude that the majority of the hitting times, $\numslots_0$, satisfy $150 \leq \numslots_0 \leq 450$.

\begin{figure}[h]
    \centerline{\includegraphics[width=8cm]{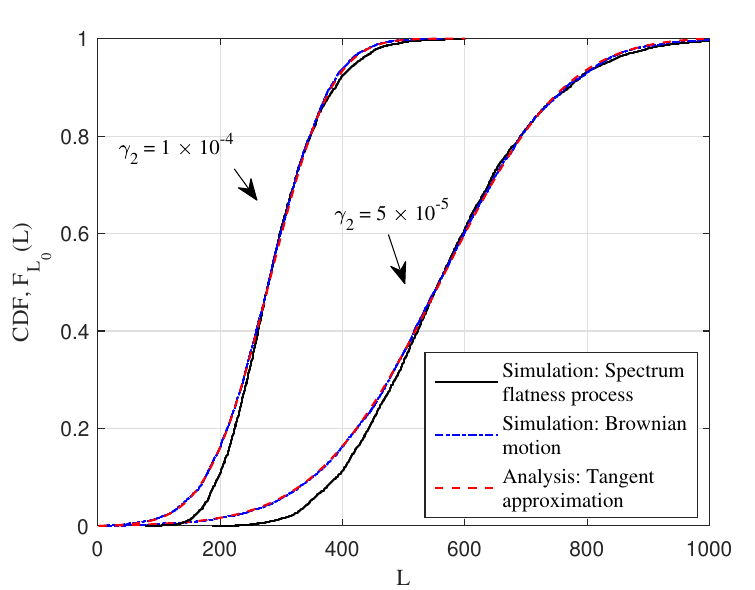}}
    \caption{The empirical CDFs of $L_0$ satisfying \eqref{eq:ch_adpt_rmsbw_req} and the Brownian motion process crossing the bound in \eqref{eq:ch_adpt_BM_rqmt}, and the corresponding tangent approximations for $\numfreqs=32$ and $\gamma_2 \in \{5\times 10^{-5},1\times10^{-4}\}$.} \label{fig:ch_adpt_CDFs_M32}
\end{figure}

To provide a detailed insight into the hitting time, Fig. \ref{fig:ch_adpt_CDFs_M32} plots the empirical CDFs of the first subpulse number, $L_0$, satisfying the spectrum flatness requirement in \eqref{eq:ch_adpt_rmsbw_req} and the Brownian motion process crossing the bound in \eqref{eq:ch_adpt_BM_rqmt}, as well as the corresponding tangent approximations for $\numfreqs=32$, $\gamma_2 \in \{5\times 10^{-5},1\times10^{-4}\}$. We observe that when $\gamma_2 = 1\times10^{-4}$, the majority of the $\numslots_0$ values satisfy $\numslots_0 \in \{150,\cdots, 450\}$, confirming the conclusion drawn from Fig. \ref{fig:ch_adpt_mean_pm_std_M32_gamma1e-4}. In addition, the Brownian model accurately approximates the spectrum flatness process for large $\numslots$ for both $\gamma_2$ values. Note that there is a deviation in the small $\numslots$ region since the Gaussian approximation to $\uspec(\numslots)$ is less accurate for small $\numslots$. However, the probability that $\numslots$ lies in this region is not significant. For example, if we expect the Brownian model to approximate the true hitting time distribution with a maximum error of $5\%$ for $\numfreqs = 32, \gamma_2 = 1\times 10^{-4}$, the value of $\numslots$ should be greater than $200$, while the probability that $\numslots<200$ is only around $0.1$. In addition, we note that the key purpose of approximating the hitting time distribution is to provide an estimate of the probability of having extremely long waveforms. Thus, the high accuracy in the upper tail is very useful, while the deviation in the lower tail is acceptable. We also observe that the tangent approximation always closely follows the Brownian model. Therefore, the theoretical analysis of the hitting time distribution derived using the Brownian model and the tangent approximation is useful.

\subsection{Radar Performance Evaluation} \label{sec:num_radar}
\begin{figure}[h]
    \centerline{\includegraphics[width=8cm]{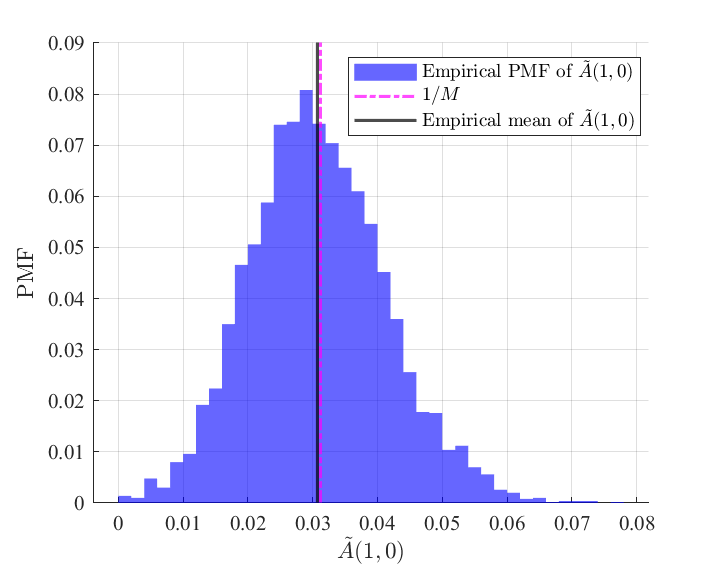}}
    \caption{The empirical PMF of the grid point AF SLs at $\tau = \pri, \omega = 0$ of all generated waveform realisations for $\numfreqs=32$ and $\gamma_2 = 1 \times 10^{-4}$.} \label{fig:ch_adpt_A10_M32_gamma1e-4}
\end{figure}

Next, we focus on other performance metrics of the radar waveform. Fig. \ref{fig:ch_adpt_A10_M32_gamma1e-4} plots the empirical probability mass function (PMF) of the grid point AF SLs at $\tau = \pri, \omega = 0$ of all generated waveform realisations for $\numfreqs=32$ and $\gamma_2 = 1 \times 10^{-4}$ as a histogram. Note that we focus on $\Tilde{A}(1,0)$ instead of the PSL since it requires extremely high computational power to compute the entire AF and find the PSL for large $\numslots$ values and all generated waveform realisations. Hence, we select the grid point AF SL with the largest mean value. We observe that the empirical PMF of $\Tilde{A}(1,0)$ concentrates around the empirical mean, whose value can be accurately approximated by $1/\numfreqs$. This result shows the theoretical analysis of $E[\Tilde{A}(1,0)]$ in \eqref{eq:AF_kr_mean_lim} for the fixed length scheme is still useful for estimating the AF SLs of the dynamic scheme. This is mainly because the subpulse number $\numslots$ satisfying the spectrum flatness requirement is usually large enough. 

\begin{figure*}[h!]
    \centerline{\includegraphics[width=17cm]{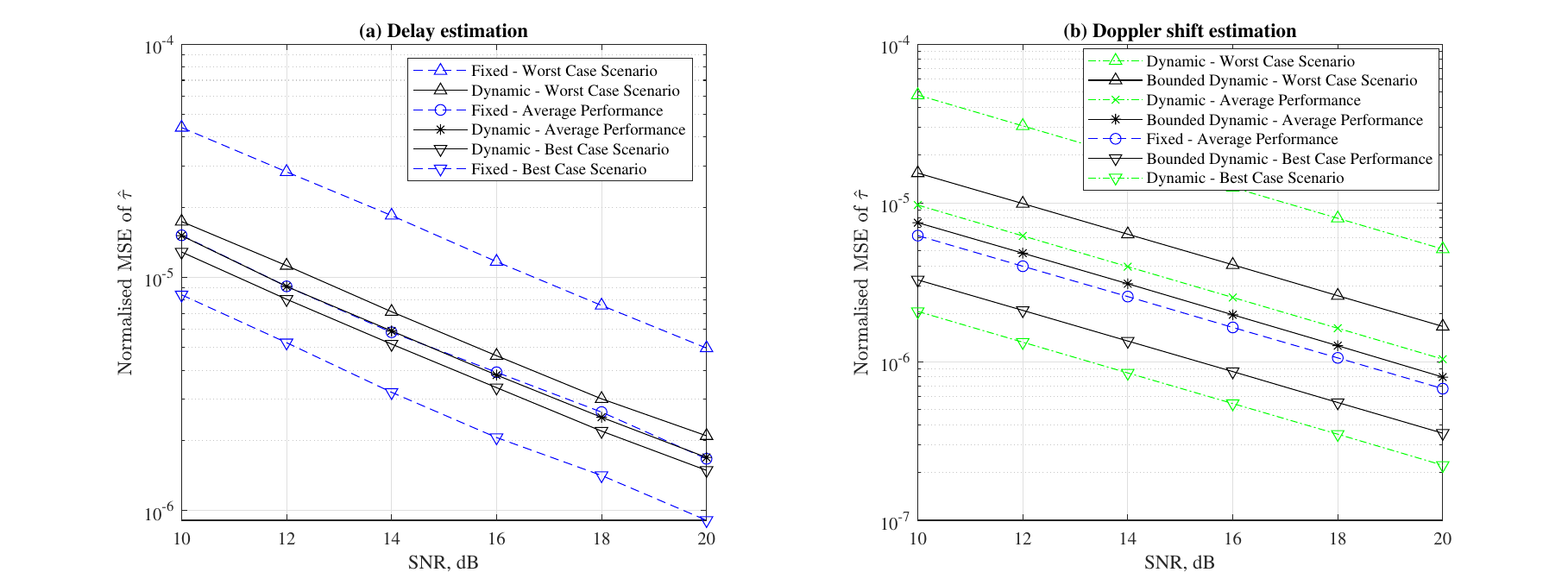}}
        \caption{The normalised (a) delay and (b) Doppler estimation MSEs for the dynamic scheme with $\gamma_2 = 1 \times 10^{-4}$ and the fixed length scheme with $\numslots = 300$. Both have $\numfreqs = 32$. Note that the number of subpulses of the fixed length scheme is selected to be close to the mean value of the dynamic scheme.} \label{fig:ch_adpt_MSE}
\end{figure*}

Fig. \ref{fig:ch_adpt_MSE} compares the delay and Doppler estimation MSEs of the proposed dynamic scheme with $\gamma_2 = 1 \times 10^{-4}$ and the traditional fixed length scheme with $\numslots = 300$. Both have $\numfreqs = 32$, ensuring that the total bandwidth occupied by both schemes is the same. Since the main focus of the numerical example section is to compare the performance of different schemes instead of discussing the impact of parameters, we choose the subpulse width $T$ to be normalised value, i.e., $1$, and the frequency separation $\Delta f$ to be $1/T = 1$. In addition, all the simulated MSE values are normalised with respect to $\pri$. Hence, if a practical $T$ value is used, the MSEs will be scaled accordingly. 
The number of subpulses of the fixed length scheme is selected to be the empirical mean subpulse number of the dynamic scheme, rounded to the closest integer, whose value is $L = 300$. All MSE performance versus SNR curves are generated using Monte Carlo simulations, where the SNR is defined as the ratio of the energy of the received signal part, $\sigma_{\Tilde{b}}^2 \int_0^{LT}|s(t)|^2 dt$, to the noise PSD, $N_0$. Note that in practical systems, longer waveforms always accumulate more energy and thus have higher SNR. Nevertheless, for the fairness of comparison, in simulations we change $N_0$ to make sure that the SNR is equivalent to particular values even with different subpulse numbers. The legends of the two subplots are explained as follows.
\begin{enumerate}
    \item ``Fixed" denotes the fixed length scheme. The fixed length scheme is simulated to generate a number of waveform realisations with the same $\numslots$.
    \begin{enumerate}
        \item[a)] The best and worst case scenarios in Fig. \ref{fig:ch_adpt_MSE} (a): Among these fixed length waveform realisations, the best and the worst case scenarios for the delay estimation focus on the ones with the largest and the smallest RMS bandwidths, respectively.
        \item[b)] Average performance in Fig. \ref{fig:ch_adpt_MSE} (a):  The averaged delay estimation MSE is averaged among all fixed length realisations.
        \item[c)] Average performance in Fig. \ref{fig:ch_adpt_MSE} (b):  The averaged Doppler estimation MSE is averaged among all fixed length realisations.
    \end{enumerate}
    \item ``Dynamic" denotes the proposed dynamic subpulse number. The dynamic scheme is simulated to generate a number of waveform realisations with variable $\numslots$'s.
    \begin{enumerate}
        \item[a)] The best and worst case scenarios in Fig. \ref{fig:ch_adpt_MSE} (a): Among these dynamic waveform realisations, the best and the worst case scenarios for the delay estimation focus on the ones with the largest and the smallest RMS bandwidths, respectively.
        \item[b)] The best and worst case scenarios in Fig. \ref{fig:ch_adpt_MSE} (b): Among these dynamic waveform realisations, the best and the worst case scenarios for the Doppler estimation focus on the ones with the largest and the smallest numbers of subpulses. 
        \item[c)] Average performance in Fig. \ref{fig:ch_adpt_MSE} (a):  The averaged delay estimation MSE is averaged among all dynamic realisations.
        \item[d)] Average performance in Fig. \ref{fig:ch_adpt_MSE} (b):  The averaged Doppler estimation MSE is averaged among all dynamic realisations.
        \item[e)] ``Bounded" in Fig. \ref{fig:ch_adpt_MSE} (b): The upper and lower bounds on the number of subpulses are introduced to control the maximal processing latency as in \eqref{eq:ch_adpt_length_req} and the Doppler estimation performance \eqref{eq:ch_adpt_rmstd_req_}, respectively. In other words, all waveform realisations with $\numslots$ not satisfying the bounds are excluded in the curves with ``Bounded" in their labels. The use of bounds will be discussed in detail below.
    \end{enumerate}
\end{enumerate}

We observe from Fig. \ref{fig:ch_adpt_MSE} (a) and (b) that the dynamic and the fixed length schemes have similar average MSE performance in terms of both the delay and Doppler estimations. This is mainly because the subpulse number of the fixed length scheme is close to the empirical mean value of the dynamic scheme. Nevertheless, the performance of the best and the worst case scenarios for the two schemes has a significant difference. Since the dynamic scheme guarantees a flat frequency spectrum, different dynamic waveform realisations have similar RMS bandwidths, resulting in a stable delay estimation MSE. In contrast, the fixed length scheme does not have such a guarantee, leading to a large variance in the delay estimation MSE, as can be visualised from Fig. \ref{fig:ch_adpt_MSE} (a). The opposite trend is seen for the Doppler estimation. Here, the MSE for the fixed length scheme is stable, while that for the dynamic scheme varies due to the variance in the subpulse number. As has been discussed in Section \ref{sec:ch_adpt_sig_mod}, in practice metrics other than the spectrum flatness can be introduced to control the performance. Following the ideas in \eqref{eq:ch_adpt_rmstd_req_} and \eqref{eq:ch_adpt_length_req}, we introduce lower and upper bounds on the subpulse number, $200 \leq \numslots \leq 400$, to the dynamic scheme. The lower bound is used to make sure that each radar waveform satisfies the minimum requirement on the Doppler estimation capability. On the other hand, the upper bound is used to control the maximum processing latency, which also controls the variability of the Doppler estimation capability. Note that the bounds have negligible impact on the delay estimation MSE since the RMS bandwidth is not directly related to the number of subpulses, as has been analysed in Appendix \ref{sec:dop_time_del_bw}. We observe from Fig. \ref{fig:ch_adpt_MSE} (b) that there is a significant improvement in the stability of the Doppler estimation MSE, which indicates the effectiveness of the bounds. We also note that the average Doppler estimation MSE is slightly decreased. Although waveform realisations with unacceptable RMS time durations are excluded by the lower bound, the probability of $\numslots<200$ or $\numslots>400$ is low, as can be shown by the CDF plots for $\gamma_2 = 1 \times 10^{-4}$ in Fig. \ref{fig:ch_adpt_CDFs_M32}. Thus, the decrease in the average Doppler estimation MSE is insignificant.

\begin{figure*}[h!]
    \centerline{\includegraphics[width=17cm]{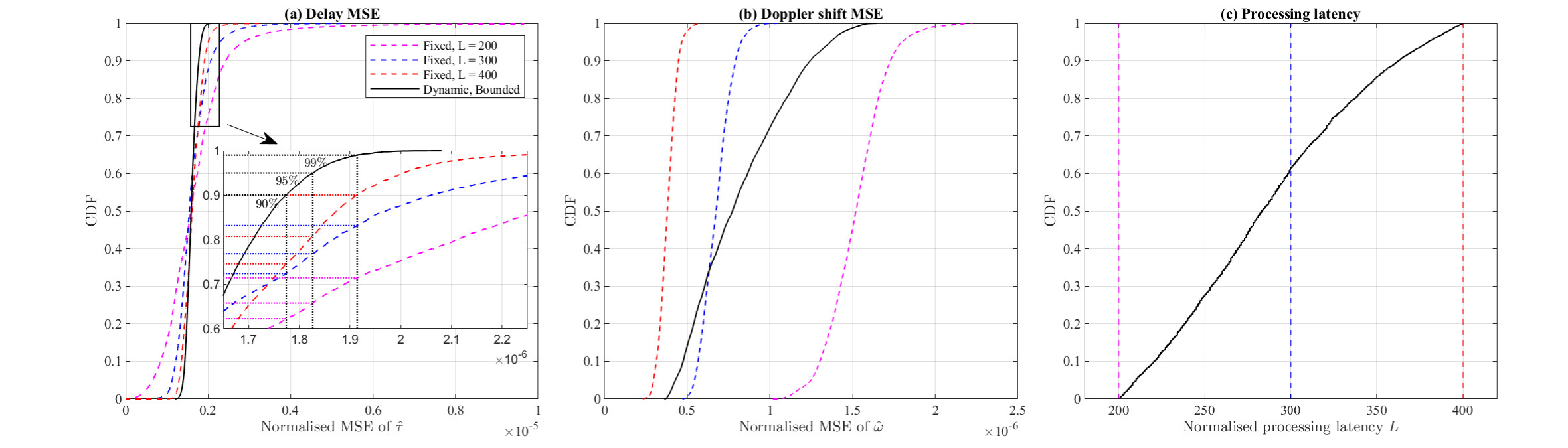}}
        \caption{The CDFs of normalised (a) delay, (b) Doppler MSEs at $20$dB SNR, and (c) the normalised processing latency for the dynamic scheme with $\gamma_2 = 1 \times 10^{-4}$ and the fixed length scheme with $\numslots \in \{200, 300, 400\}$. All have $\numfreqs = 32$. In (a), the three intersections, formed by pairs of perpendicular black dotted lines, present $90\%$, $95\%$ and $99\%$ in the dynamic waveform delay MSE CDF upper tail, respectively, each corresponds to a delay MSE value.} 
        \label{fig:Comp_SNR20}
\end{figure*}

Fig. \ref{fig:Comp_SNR20} provides a comprehensive comparison between the ``bounded" dynamic scheme and various fixed length schemes using empirical CDFs of (a) delay and (b) Doppler estimation MSEs at a particular SNR value, $20$dB, and (c) the processing latency caused by the generation of each waveform. The parameters are the same as in Fig. \ref{fig:ch_adpt_MSE}, except that we consider $\numslots \in \{200, 300, 400\}$ for the fixed length scheme. Note that the number of subpulses of the dynamic scheme satisfies $200 \leq \numslots \leq 400$. Hence, the three $\numslots$ values for the fixed length scheme correspond to the lower bound, the mean and the upper bound of the subpulse number of the dynamic scheme, respectively. The CDFs are generated using thousands of waveform realisations, with the performance of each generated based on Monte Carlo simulations. 
We observe from Fig. \ref{fig:Comp_SNR20} (a) that the dynamic scheme achieves better stability of the delay estimation capability compared to the fixed length schemes, as expected. The variation in the delay estimation capability of the fixed length schemes can be visualised by the long upper tails of the CDFs, especially for small $\numslots$ values. The delay MSE stability comparison between the dynamic and the fixed length schemes can also be made from a perspective of a probabilistic performance guarantee, which can be visualised by zooming into a part of the upper tails as in Fig. \ref{fig:Comp_SNR20} (a). For example, the $95\%$ point of the dynamic scheme is only attained $76.84\%$ of the time when $L=300$. For completeness, Table \ref{tab:CDFvals} lists all delay MSE CDF values for fixed length schemes corresponding to $90\%$, $95\%$ and $99\%$ in the CDF for the dynamic scheme. The upper tail performance indicates that with a particular spectrum control for each single waveform as in \eqref{eq:ch_adpt_rmsbw_req}, the dynamic scheme is superior to the fixed length ones in terms of the delay estimation performance stability.

On the other hand, Fig. \ref{fig:Comp_SNR20} (b) indicates that the Doppler estimation capability of the dynamic scheme is more variable than fixed length schemes. Nevertheless, it is controlled by the bounds on the subpulse number, as its values do not exceed those of fixed length schemes with $\numslots = 200$ and $\numslots = 400$. Note that although all FSK waveforms with the same $L$ value should have the same Doppler estimation capability as in Appendix \ref{sec:dop_time_del_bw}, the variation within each fixed length scheme is unavoidable due to the random noise and reflection complex amplitude realisations. 
The processing latency caused by the generation of each waveform in Fig. \ref{fig:Comp_SNR20} is represented by the number of subpulses, $\numslots$, of each waveform realisation. Hence, the dynamic scheme results in a variable processing latency, whose value is between the two bounds. For the ease of comparison, Table \ref{tab:comp} lists the qualitative characteristics of all schemes. It is worth noting that there is always a tradeoff between the Doppler estimation capability and the processing latency as increasing $L$ leads to improved Doppler MSE but longer processing latency. While this unavoidable tradeoff exists no matter which scheme is used, the proposed scheme always achieves a highly stable delay estimation capability as expected.

In practical radar systems, a single waveform usually has an insufficient pulse width, resulting in an undesirable Doppler estimation capability \cite[Chapter 3]{richards14}. A standard approach to improve the Doppler estimation capability is to use a coherent train of multiple waveforms, enlarging the ``pulse width" to an entire coherent processing interval (CPI). The delay estimation is performed in fast time, i.e., within each single waveform, while the Doppler estimation is performed in slow time, i.e., across multiple waveforms. Therefore, the Doppler estimation capability of a single waveform is not as critical as its delay estimation capability. This makes the proposed dynamic scheme highly attractive in practical multi-waveform radar processing scenarios, given that it provides a superior delay estimation stability compared to its fixed length counterparts.

\begin{table*}[!ht] 
\caption{Delay MSE CDF values for fixed length schemes corresponding to particular CDF values for the dynamic scheme in Fig. \ref{fig:Comp_SNR20} (a).} 
\vspace{-1.2\baselineskip}
\begin{center}
\renewcommand{\arraystretch}{1.1}
\begin{tabular}{c | c | c | c }
\hline\hline
Dynamic, ``bounded" & $90\%$ & $95\%$ & $99\%$	\\
\hline
Fixed length, $\numslots = 200$ & $62.23\%$	& $65.74\%$ & $71.38\%$ \\
\hline
Fixed length, $\numslots = 300$ & $72.33\%$	& $76.84\%$ & $83.17\%$  	\\
\hline
Fixed length, $\numslots = 400$ & $74.53\%$	& $80.73\%$ & $90.03\%$  	\\
\hline \hline
\end{tabular}
\label{tab:CDFvals}  
\end{center}
\end{table*}

\begin{table*}[!ht] 
\caption{The qualitative comparison between the dynamic scheme and fixed length schemes in Fig. \ref{fig:Comp_SNR20}.} 
\vspace{-1.2\baselineskip}
\begin{center}
\renewcommand{\arraystretch}{1.1}
\begin{tabular}{c | c | c | c }
\hline\hline
Scheme & Delay estimation & Doppler estimation & processing latency \\ 
\hline
Dynamic, ``bounded" & The most stable & Unstable but controlled by bounds & Unstable but controlled by bounds	\\
\hline
Fixed length, $\numslots = 200$ & \makecell{ The most unstable, realisations \\ with very large MSEs exist}    & \makecell{ The largest Doppler MSE on average, \\ with the second largest variance} 	& \makecell{Constant, with the shortest\\ processing latency}	 \\
\hline
Fixed length, $\numslots = 300$  & \makecell{ The second most unstable, \\ realisations with large MSEs exist}	& \makecell{ Similar Doppler MSE as the dynamic \\ scheme, with a smaller variance}	& \makecell{Constant, with a moderate\\ processing latency}	\\
\hline
Fixed length, $\numslots = 400$  & \makecell{ Not as stable as the dynamic scheme, \\ better than schemes with smaller $\numslots$ values}	& \makecell{ The smallest Doppler MSE on average, \\ with the smallest variance} & \makecell{Constant, with the longest\\ processing latency}	\\
\hline \hline
\end{tabular}
\label{tab:comp}  
\end{center}
\end{table*}

\section{Conclusions and Future Extensions} \label{sec:conclusion}
In this paper, we propose an FSK-based dynamic subpulse number JCR waveform design. The transmitter continuously generates and sends subpulses based on $\numfreqs$-ary FSK for communications. Generation continues until the spectrum of the entire waveform is sufficiently uniform across the whole available bandwidth and deterministic limits to control AF SLs, RMS time duration, and overall length are satisfied. The waveform up to this moment is regarded as the radar sensing waveform and is used for delay and Dopper estimation. This guarantees the estimation performance of each processed radar waveform. Note that the scheme generates the most reliable waveforms if the random length prescribed by the spectrum flatness metric falls within the deterministic limits given by the other metrics. In this situation, all the radar metrics are simultaneously satisfied. In order to assess the effectiveness of the scheme, we therefore analyse the probability distribution of the number of subpulses. Since the spectrum flatness is monitored once each subpulse is generated, the problem is the distribution of the hitting time of a stochastic process, which is often a complex or intractable problem. Nevertheless, using the statistical properties of the spectrum flatness process, we developed a novel approximation of the process as a function of Brownian motion. An existing solution for the one-sided Brownian exit density, namely, the tangent approximation, is applied to provide a theoretical analysis of the hitting time distribution of the spectrum flatness process. Numerical examples are provided to evaluate the accuracy of the approximate distribution. The AF PSL and the delay and Doppler shift estimation performance of the radar waveforms are also evaluated using numerical examples.

Possible future extensions of the existing work include the following:
\begin{enumerate}
    \item We can observe from Fig. \ref{fig:ch_adpt_CDFs_M32} that the Brownian model cannot accurately approximate the actual subpulse number distribution when $\numslots$ is small. Although it is acceptable due to the purpose of the approximation and the low probability of $\numslots$ falling into the ``inaccurate" region, it would be useful to improve the accuracy of the theoretical analysis in this region.
    
    \item Although numerical examples show that the delay estimation MSE is stable and controllable, some other performance metrics can be considered to further enhance the performance of radar waveforms. In this situation, the analysis of the subpulse distribution will be useful to assess which performance metrics deliver the best results.

    \item 
    While the existing work considers constant modulus waveforms only, sometimes a low PAPR value, instead of being strictly one, would be enough. This would allow for the use of more popular multi-carrier waveforms. More specifically, the OFDM index modulation (IM) based scheme \cite{Tsonev11}, where a portion of subcarriers are silent, is a potential candidate to control the PAPR level. By incorporating the idea of dynamic waveforms into the OFDM-IM scheme, the designed scheme would be able to satisfy the PAPR constraint and the communications and/or radar performance requirements.
\end{enumerate}

\begin{appendices}
\section{The Advantages of Long FSK Waveforms and Performance Metrics} \label{sec:ch_adpt_large_no_subpulses}

\subsection{AF Sidelobe Levels}  \label{sec:clutter_afsl}



The AF SL of an FSK-modulated waveform at the delay and Doppler pair $(\tau,\omega) = (k \pri, 2 \pi r \freqsep), (k,r) \in \mathcal{D} = \{(k,r)| k \in \{0, \cdots, \numslots - 1\}, r \in \{-(\numfreqs-1), \cdots, \numfreqs - 1\}\} \setminus \{(k,r)| k \leq 0, r = 0 \}$ can be expressed as \cite[eq.(16)]{Han25global}
\begin{align}\label{eq:AF_FSK_grid}
\Tilde{A}(k,r) = \Bigg| \frac{1}{\numslots}\sum_{l = k}^{\numslots - 1}X_{k,r}(l) \Bigg| , 
\end{align}
where $X_{k,r}(l)$
\begin{align}\label{eq:Xkrl}
X_{k,r}(l) = \left\{
\begin{array}{ll} 1~~~~~~~~\omega_{l-k}-\omega_l = 2\pi r \freqsep\\
0~~~~~~~~\omega_{l-k}-\omega_l \neq 2\pi r \freqsep.
\end{array}\right.
\end{align}
It has been shown in \cite{Han25global} that $\Tilde{A}(k,r)$ follows a scaled binomial distribution, i.e., $\numslots \times \Tilde{A}(k,r) \sim \text{B}(\numslots-k, (\numfreqs-|r|)/\numfreqs^2)$. Using the binomial mean and variance equations, the mean and the variance of $\Tilde{A}(k,r)$ can be expressed as 
\begin{align} \label{eq:AF_kr_mean}
    E\left[\Tilde{A}(k,r)\right] = \frac{\numslots-k}{\numslots}\frac{\numfreqs-|r|}{\numfreqs^2},
\end{align}
\begin{align}  \label{eq:AF_kr_var}
    \text{Var}\left[\Tilde{A}(k,r)\right] = \frac{\numslots-k}{\numslots^2}\frac{(\numfreqs-|r|)(\numfreqs^2-\numfreqs+|r|)}{\numfreqs^4}.
\end{align}
We note from \eqref{eq:AF_kr_var} that the variance decreases with $\numslots$. Hence, the SL stabilises at its mean in \eqref{eq:AF_kr_mean} with an increase in $\numslots$. In addition, we observe from \eqref{eq:AF_kr_mean} that the mean of a SL closer to the origin is larger, for fixed waveform parameters $\numslots$ and $\numfreqs$. Since SLs stabilise at their mean values, the peak SL (PSL) tends to appear at $(\tau,\omega) = (k \pri, 2 \pi r \freqsep)$ with small $k$ and $|r|$ when $\numslots$ is large. 

More specifically, for particular small $k$ and $|r|$ values, we can easily obtain 
\begin{align} \label{eq:AF_kr_mean_lim}
    \lim_{\numslots \rightarrow \infty}\left(E\left[\Tilde{A}(k,r)\right]\right) = \frac{\numfreqs-|r|}{\numfreqs^2},
\end{align}
\begin{align}  \label{eq:AF_kr_var_lim}
    \lim_{\numslots \rightarrow \infty}\left(\text{Var}\left[\Tilde{A}(k,r)\right]\right) = 0.
\end{align}
We observe from \eqref{eq:AF_kr_mean_lim} and \eqref{eq:AF_kr_var_lim} that AF SLs at small delay and Doppler values stabilise to $1/\numfreqs$ for large $\numslots$. In other words, when the radar waveform is long enough, we can control the AF SLs and the PSL to an acceptable level by designing $\numfreqs$.

\subsection{Delay-Doppler Estimation Capability and RMS time-bandwidth properties} \label{sec:dop_time_del_bw}


It has been shown in \cite[eq.(10.94)-(10.95)]{vantrees01} that the CRLBs on the Doppler and delay estimation error variances, $\text{CRLB}_{\omega}$ and $\text{CRLB}_{\tau}$, can be expressed as
\begin{align} 
    \text{CRLB}_{\omega} = \frac{1}{C\sigma_t^2},  \label{eq:crlbw}
\end{align}
\begin{align} 
    \text{CRLB}_{\tau} = \frac{1}{C\sigma_{\omega}^2},  \label{eq:crlbt}
\end{align}
where $C$ is related to the received SNR. The denominators, $\sigma_t^2$ and $\sigma_{\omega}^2$, are the squared RMS time duration and RMS bandwidth of the waveform, respectively, emphasising the importance of the time-bandwidth product of a radar waveform \cite[eq.(10.63), eq.(10.65)]{vantrees01}. They are also proportional to the absolute curvature of the AF at the origin along the Doppler and the delay axes, respectively, showing their relationships with the AF mainlobe \cite[eq.(10.96), eq.(10.98)]{vantrees01}.

The RMS time duration of an FSK modulated JCR waveform can be expressed as \cite[eq.(14)]{Han23local}
\begin{align}
    \sigma_{\tau}^2 = \numslots^2 \pri^2/12. \label{eq:secdop_J22}
\end{align}
It is obvious that the RMS bandwidth in \eqref{eq:secdop_J22} increases with $\numslots$, indicating that the Doppler shift estimation capability improves with $\numslots$.

The RMS bandwidth of an FSK modulated JCR waveform can be expressed as \cite[eq.(22)]{Han23local}
\begin{align} \label{eq:secdel_J11}   
\sigma_{\omega}^2 & \approx \frac{(\pi \bw)^2}{\pi \bw \pri \text{Si}(\pi \bw \pri) + \cos(\pi \bw \pri) - 1} \\ \notag
& + \left(\frac{2\pi \freqsep}{\numslots}\right)^2 \sum_{m=0}^{\numfreqs-1} \sum_{n=m+1}^{\numfreqs-1}N_m N_n (m-n)^2,   
\end{align}
where $N_m$ is the number of times the $m$-th frequency, $2\pi(f_0+(m-1)\Delta f)$, appears in the sequence of $\numslots$ frequencies and $\bw$ is the limiting bandwidth applied to the rectangular pulse shaping function. Note that $\bw$ is introduced since the integrals the RMS bandwidth calculation do not converge when the perfect rectangular pulse is used. Nevertheless, \cite{Han23local} has shown that $\bw$ has an insignificant impact on the RMS bandwidth since the second term of \eqref{eq:secdel_J11} is dominant. 
The variance of $\sigma_{\omega}^2$ can be expressed as \cite[eq.(27)]{Han23local}
\begin{align} \label{eq:secdel_var_J11}
\text{Var}\left[\sigma_{\omega}^2\right] \approx & (2\pi \freqsep)^4( \numslots - 1)( \numfreqs+1)( \numfreqs-1) \\ \notag 
&\times (2 \numslots \numfreqs^2 - 8 \numslots + 3 \numfreqs^2 + 3)/(360\numslots^3). 
\end{align}
It is obvious that $\text{Var}\left[\sigma_{\omega}^2\right]$ decreases with $\numslots$, indicating that the RMS bandwidth tends to stabilise at its mean value \cite[eq.(26)]{Han23local},
\begin{align} \label{eq:secdel_E_J11}
E\left[\sigma_{\omega}^2\right] \approx & \frac{(\pi \bw)^2}{\pi \bw \pri \text{Si}(\pi \bw \pri) + \cos(\pi \bw \pri) - 1}  \\ \notag 
& + \frac{(2\pi \freqsep)^2(\numslots -1 )(\numfreqs+1)(\numfreqs-1)}{12\numslots}, 
\end{align}
with an increase in $\numslots$. Although $E[\sigma_{\omega}^2]$ depends on $\numslots$, when $\numslots \rightarrow \infty$ we obtain  
\begin{align}  \label{eq:ch_adpt_E_sigmaw2}
    \lim_{\numslots \rightarrow \infty} E[\sigma_{\omega}^2] \approx & \frac{(\pi \bw)^2}{\pi \bw \pri \text{Si}(\pi \bw \pri) + \cos(\pi \bw \pri) - 1}   \\  \notag
    & + \frac{(2\pi \freqsep)^2(\numfreqs+1)(\numfreqs-1)}{12}. 
\end{align}
We observe from \eqref{eq:ch_adpt_E_sigmaw2} that $E[\sigma_{\omega}^2]$ coincides with the value of $\sigma_{\omega}^2$ for a waveform whose centre frequencies of subpulses are evenly spread among the available bandwidth when $\numslots > \numfreqs$. More specifically, by assuming that $\numslots/\numfreqs$ is an integer and substituting $N_m = \numslots/\numfreqs, \forall m \in\{0, \cdots, \numfreqs-1\}$ into \eqref{eq:secdel_J11}, we obtain
\begin{align} \label{eq:ch_adpt_sigmaw2_u}
\left(\sigma^2_{\omega}\right)_{u} \approx & \frac{(\pi \bw)^2}{\pi \bw \pri \text{Si}(\pi \bw \pri) + \cos(\pi \bw \pri) - 1}   \\  \notag
    & + \frac{(2\pi \freqsep)^2(\numfreqs+1)(\numfreqs-1)}{12}, 
\end{align}
where the subscript $u$ is introduced to denote that the frequency spectrum of the waveform is uniform. Note that $N_m$ is the number of times that $2\pi(f_0+m\Delta f)$ appears in the frequency sequence. Combining the result of vanishing $\text{Var}[\sigma_{\omega}^2]$ in \eqref{eq:secdel_var_J11} and \eqref{eq:ch_adpt_E_sigmaw2}, we conclude that the RMS bandwidth stabilises at the value for evenly distributed centre frequencies with a decreasing variation as $\numslots$ increases. Therefore, $\sigma_{\omega}^2$ becomes very predictable for large $\numslots$. More specifically, since the corresponding RMS bandwidth in \eqref{eq:ch_adpt_E_sigmaw2} is $\mathcal{O}(\numfreqs^2)$, we can guarantee an acceptable delay estimation capability of a waveform with large $\numslots$ by designing $\numfreqs$.

{\color{black}\subsection{Frequency Spectrum Flatness} \label{sec:metric_spec_flat}
As has been discussed in Appendix \ref{sec:dop_time_del_bw}, a waveform with a large RMS bandwidth has a better delay estimation accuracy. 
Comparing \eqref{eq:ch_adpt_E_sigmaw2} with \eqref{eq:ch_adpt_sigmaw2_u}, we observe that a large enough $\numslots$ value can guarantee stabilisation towards a situation in which the frequencies are almost evenly distributed among the available spectrum. A flat spectrum occupying the whole available bandwidth, which is common for traditional stepped-frequency radar waveforms, is beneficial since it represents an RMS bandwidth comparable to traditional stepped-frequency radar waveforms. This stable RMS bandwidth value can be controlled by designing waveform parameters such as $\numfreqs$, as is shown in \eqref{eq:ch_adpt_E_sigmaw2}. 
The flatness of the frequency spectrum of an FSK waveform can be assessed by calculating a mean squared error (MSE), which quantifies the deviation of the observed frequency distribution from the desired flat spectrum, i.e., 
\begin{align} \label{eq:spctrm_fltns_mtrc}
    \uspec = \frac{\sum_{m=0}^{\numfreqs-1}\left(\frac{N_m(\numslots)}{L}-\frac{1}{\numfreqs}\right)^2}{\numfreqs},
\end{align}
where 
$N_m$ denotes the number of the times the $m$-th frequency was selected in the frequency sequence. The smaller the value of $U$, the flatter the frequency spectrum is. When $\uspec=0$, $N_m(\numslots) = \numslots/\numfreqs, \forall m\in\{0,\cdots, \numfreqs-1\}$, indicating a perfect evenly distributed frequency spectrum across the entire available bandwidth. }

\section{The Derivations of the Brownian Approximation}
\subsection{The Statistics of $\chi^2(L)$} \label{app:autostats_chi}
The mean and the variance of the chi-squared test statistic, $\chi^2(\numslots)$, can be expressed as \cite{Lawal1980}
\begin{align}   
    E[\chi^2(\numslots)] &= \numfreqs - 1, \label{eq:ch_adpt_E_chisq} \\
    \text{Var}[\chi^2(\numslots)] &= 2(\numfreqs-1)\left(1-\frac{1}{\numslots}\right). \label{eq:ch_adpt_var_chisq}
\end{align}
The correlation between $\chi^2(\numslots)$ and $\chi^2(\numslots+k)$ is defined as
\begin{align}  \label{eq:ch_adpt_cor_chisq_1}
    &\rho[\chi^2(\numslots), \chi^2(\numslots+k)] \\ \notag
    &= \frac{E[\chi^2(\numslots)\chi^2(\numslots+k)] - E[\chi^2(\numslots)]E[\chi^2(\numslots+k)]}{\sqrt{\text{Var}[\chi^2(\numslots)]\text{Var}[\chi^2(\numslots+k)]}},
\end{align}
where $E[\chi^2(\numslots)\chi^2(\numslots+k)]$ can be calculated as \eqref{eq:ch_adpt_E_L_Lpk}. The equality $\eqsymbol{(a)}$ in \eqref{eq:ch_adpt_E_L_Lpk} is achieved by substituting $\chi^2(\numslots+k)$ into \eqref{eq:ch_adpt_E_L_Lpk} using \eqref{eq:ch_adpt_chisq}. The equality $\eqsymbol{(b)}$ is based on the obvious fact that the $k$ extra subpulses are generated independent of the previous $\numslots$ subpulses based on $\numfreqs$-ary FSK. Hence, $(N_m(\numslots+k)-N_m(\numslots)), m \in \{0, \cdots, \numfreqs-1\}$ are independent of $N_m(\numslots), m \in \{0, \cdots, \numfreqs-1\}$, following a multinomial distribution with the probabilities $\Pr[\omega_m] = 1/\numfreqs, m \in \{0, \cdots, \numfreqs-1\}$ and the number of trials $\sum_{m=0}^{\numfreqs-1}(N_m(\numslots+k)-N_m(\numslots))=k$. Thus, $E[N_m(\numslots+k)-N_m(\numslots)] = k/\numfreqs, m \in \{0, \cdots, \numfreqs-1\}$. The equality $\eqsymbol{(c)}$ is then achieved by substituting $E[\chi^2(\numslots)]$, $\text{Var}[\chi^2(\numslots)]$ and $E[\chi^2(k)]$ into \eqref{eq:ch_adpt_E_L_Lpk} using \eqref{eq:ch_adpt_E_chisq} and \eqref{eq:ch_adpt_var_chisq}. 
\begin{figure*}[!t]
\begin{align}   \label{eq:ch_adpt_E_L_Lpk}
& E[\chi^2(\numslots)\chi^2(\numslots+k)] \eqsymbol{(a)} E\left[\chi^2(\numslots) \frac{\sum_{m=0}^{\numfreqs-1}\left(N_m(\numslots+k)-\frac{\numslots+k}{\numfreqs}\right)^2}{\frac{\numslots+k}{\numfreqs}}\right]   \notag \\
& = E\left[ \chi^2(\numslots) \frac{\sum_{m=0}^{\numfreqs-1}\left(N_m(\numslots)-\frac{\numslots}{\numfreqs} + (N_m(\numslots+k)-N_m(\numslots))-\frac{k}{\numfreqs}\right)^2}{\frac{\numslots}{\numfreqs}} \frac{\numslots}{\numslots+k}\right]   \notag \\
& = E \Bigg[ \chi^2(\numslots) \Bigg(\chi^2(\numslots) + \frac{2\sum_{m=0}^{\numfreqs-1}\left(N_m(\numslots)-\frac{\numslots}{\numfreqs}\right)\left((N_m(\numslots+k)-N_m(\numslots))-\frac{k}{\numfreqs}\right)}{\frac{\numslots}{\numfreqs}} + \frac{\sum_{m=0}^{\numfreqs-1}\left((N_m(\numslots+k)-N_m(\numslots))-\frac{k}{\numfreqs}\right)^2}{\frac{\numslots}{\numfreqs}}\Bigg) \frac{\numslots}{\numslots+k} \Bigg]   \notag \\
& \eqsymbol{(b)} \frac{\numslots}{\numslots+k} E\left[\left( \chi^2(\numslots) \right)^2\right] + \frac{k}{\numslots+k} E\left[\chi^2(\numslots)\right] E\left[\frac{\sum_{m=0}^{\numfreqs-1}\left((N_m(\numslots+k)-N_m(\numslots)) - \frac{k}{\numfreqs} \right)^2}{\frac{k}{\numfreqs}}\right] \notag \\
& \eqsymbol{(c)} \frac{\numslots}{\numslots+k}\left((\numfreqs-1)^2+2(\numfreqs-1)\left(1-\frac{1}{\numslots}\right)\right) + \frac{k}{\numslots+k}(\numfreqs-1)^2. 
\end{align}
\hrule
\end{figure*}
Substituting \eqref{eq:ch_adpt_E_chisq}, \eqref{eq:ch_adpt_var_chisq} and \eqref{eq:ch_adpt_E_L_Lpk} into \eqref{eq:ch_adpt_cor_chisq_1} and following straightforward mathematical manipulations we obtain
\begin{align}   \label{eq:ch_adpt_cor_chisq_2}
\rho[\chi^2(\numslots), \chi^2(\numslots+k)] = \frac{\numslots-1}{\numslots+k} \left(1-\frac{1}{\numslots}\right)^{-1/2} \left(1-\frac{1}{\numslots+k}\right)^{-1/2}.
\end{align}
The square root terms in \eqref{eq:ch_adpt_cor_chisq_2} make it difficult to link $\chi^2(\numslots)$ with a Brownian motion process since the correlation is complicated. To simplify it, we use the binomial expansions of $\left(1-1/\numslots\right)^{-1/2}$ and $\left(1-1/(\numslots+k)\right)^{-1/2}$, e.g.,
\begin{align}
\left(1-\frac{1}{\numslots}\right)^{-1/2} = 1+\frac{1}{2\numslots} + \mathcal{O}\left(\frac{1}{\numslots^3}\right),
\end{align}
and re-express \eqref{eq:ch_adpt_cor_chisq_2} as \eqref{eq:ch_adpt_cor_chisq}. The approximation in \eqref{eq:ch_adpt_cor_chisq} is best for large $\numslots$. 
\begin{figure*}[!t]
\begin{align}   \label{eq:ch_adpt_cor_chisq}
\rho[\chi^2(\numslots), \chi^2(\numslots+k)] &= \frac{\numslots-1}{\numslots+k} \left(1+\frac{1}{2\numslots}+\mathcal{O}\left(\frac{1}{\numslots^3}\right)\right) \left(1+\frac{1}{2(\numslots+k)}+\mathcal{O}\left(\frac{1}{\numslots^3}\right)\right) \notag \\
&= \frac{1}{\numslots+k}\left(\numslots-1 + \frac{1}{2}\left(1-\frac{1}{\numslots}\right) + \frac{\numslots-1}{2(\numslots+k)} + \mathcal{O}\left(\frac{1}{\numslots^2}\right)\right) \notag \\
& = \frac{\numslots}{\numslots+k} + \mathcal{O}\left(\frac{k}{\numslots^2}\right) \notag \\
& \approx \frac{\numslots}{\numslots+k}.
\end{align}
\hrule
\end{figure*}

\subsection{Proof of Result \ref{result:ch2_BM}} \label{app:proof_result}
\begin{proof}
 The probability distribution of $\chi_{\text{BM}}^2(t)$ can be derived using \eqref{eq:ch_adpt_chisq_BM} and \eqref{eq:ch_adpt_BM_dist} as
\begin{align}    \label{eq:ch_adpt_chisqBM_dist}
    \chi_{\text{BM}}^2(t) \sim \mathcal{N}\left(\numfreqs-1, 2(\numfreqs-1)\left(1-\frac{1}{t}\right)\right).
\end{align}
The correlation between $\chi_{\text{BM}}^2(t)$ and $\chi_{\text{BM}}^2(t+t')$ for $t' > 0$ can be derived as \eqref{eq:ch_adpt_cor_chisqBM}, where the equality $\eqsymbol{(a)}$ is achieved by substituting \eqref{eq:ch_adpt_chisq_BM}, \eqref{eq:ch_adpt_BM_dist} and \eqref{eq:ch_adpt_chisqBM_dist} into \eqref{eq:ch_adpt_cor_chisqBM}, while the equality $\eqsymbol{(b)}$ uses \eqref{eq:ch_adpt_BM_corr}. 
Comparing \eqref{eq:ch_adpt_chisq_ctns_dist} with \eqref{eq:ch_adpt_chisqBM_dist} and \eqref{eq:ch_adpt_chisq_ctns_corr} with \eqref{eq:ch_adpt_cor_chisqBM}, we observe that $\chi^2(t)$ and $\chi_{\text{BM}}^2(t)$ have the same distribution and autocorrelation.
\end{proof}

\begin{figure*}[t!]
 \begin{align} \label{eq:ch_adpt_cor_chisqBM}
    \rho[\chi_{\text{BM}}^2(t), \chi_{\text{BM}}^2(t+t')] &= \frac{E[\chi_{\text{BM}}^2(t)\chi_{\text{BM}}^2(t+t')] - E[\chi_{\text{BM}}^2(t)]E[\chi_{\text{BM}}^2(t+t')]}{\sqrt{\text{Var}[\chi_{\text{BM}}^2(t)]\text{Var}[\chi_{\text{BM}}^2(t+t')]}} \notag \\
    &\eqsymbol{(a)}  \frac{E\left[(\numfreqs-1)^2 + 2(\numfreqs-1)\sqrt{\left(1-\frac{1}{t}\right)\left(1-\frac{1}{t+t'}\right)}\frac{W(t^2)W((t+t')^2)}{t(t+t')} \right] - (\numfreqs-1)^2}{2(\numfreqs-1)\sqrt{\left(1-\frac{1}{t}\right)\left(1-\frac{1}{t+t'}\right)}} \notag\\
    &= \frac{E\left[W(t^2)W((t+t')^2) \right]}{t(t+t')} \notag\\
    &\eqsymbol{(b)} \frac{t}{t+t'}.
\end{align}
\hrule
\end{figure*}

\subsection{The Derivation of the Hitting Time PDF} \label{app:hittime_pdf}
To apply the tangent approximation in \eqref{eq:ch_adpt_tgt_aprx} to the hitting time of \eqref{eq:ch_adpt_rmsbw_req}, we first derive the boundary $b(t)$. Since $\chi^2(\numslots) = \numslots \numfreqs^2 \uspec(\numslots)$, \eqref{eq:ch_adpt_rmsbw_req} can be re-expressed using the continuous-time approximation as
\begin{align} \label{eq:ch_adpt_chisq_rqmt}
\chi^2(t) \leq \gamma_2 \numfreqs^2 t.
\end{align}
Invoking the Brownian approximation, i.e., $\chi^2(t) \approx \chi_{\text{BM}}^2(t)$, we obtain the approximation
\begin{align} \label{eq:ch_adpt_chisq_rqmt_BM}
\chi_{\text{BM}}^2(t) \approx \chi^2(t) \leq \gamma_2 \numfreqs^2 t.
\end{align}
Substituting \eqref{eq:ch_adpt_chisq_BM} into \eqref{eq:ch_adpt_chisq_rqmt_BM} and using some straightforward mathematical manipulations, we re-write \eqref{eq:ch_adpt_chisq_rqmt_BM} as 
\begin{align} \label{eq:ch_adpt_BMt2_rqmt}
W(t^2) \leq c(t) = \frac{\gamma_2 \numfreqs^2 t^2 - (\numfreqs-1)t}{\sqrt{2(\numfreqs-1)\left(1-\frac{1}{t}\right)}}.
\end{align}
Substituting $t^2$ with $x$ and letting 
$b(x) = c(t^2)$, we re-write \eqref{eq:ch_adpt_BMt2_rqmt} as
\begin{align} \label{eq:ch_adpt_BM_rqmt}
W(x) \leq b(x) = \frac{\gamma_2 \numfreqs^2 x - (\numfreqs-1)\sqrt{x}}{\sqrt{2(\numfreqs-1)\left(1-\frac{1}{\sqrt{x}}\right)}}.
\end{align}
The first order derivative of $b(x)$ can be expressed as
\begin{align} \label{eq:ch_adpt_gamma_drvtv}
\frac{db(x)}{dx} = \frac{\gamma_2\numfreqs^2 x- \left(\frac{5}{4\gamma_2\numfreqs^2} + \frac{1}{2(\numfreqs-1)}\right)\sqrt{x}+\frac{3(\numfreqs-1)}{4}}{\sqrt{2(\numfreqs-1)}\left(1-\frac{1}{\sqrt{x}}\right)^{\frac{3}{2}}x}.
\end{align}
The tangent approximation to the PDF of the time, $x_0$, that the Brownian motion process crosses $b(x)$ can then be obtained by substituting \eqref{eq:ch_adpt_BM_rqmt} and \eqref{eq:ch_adpt_gamma_drvtv} into \eqref{eq:ch_adpt_tgt_aprx} as
\begin{align}  \label{eq:ch_adpt_t2_tgt_aprx}
f{x_0}(x) &\approx \frac{1}{\sqrt{2\pi x}} \frac{\left(\frac{\gamma_2\numfreqs^2}{2}-(\numfreqs-1)\right)\sqrt{x}+\frac{\numfreqs-1}{2}}{2\sqrt{2(\numfreqs-1)}\left(1-\frac{1}{\sqrt{x}}\right)^{3/2}x} \notag \\ 
&\exp\left(-\frac{(\gamma_2\numfreqs^2\sqrt{x}-(\numfreqs-1))^2}{4(\numfreqs-1)\left(1-\frac{1}{\sqrt{x}}\right)}\right), \quad x > 0.
\end{align}
Since $x = t^2$, the PDF of the time $t_0$, that the process $U(t)$ satisfies \eqref{eq:ch_adpt_rmsbw_req} can be approximated using \eqref{eq:ch_adpt_t2_tgt_aprx} and the transformation between random variables $t_0=\sqrt{x_0}$ and $x_0$ as
\begin{align}   \label{eq:app_t_tgt_aprx}
f_{t_0}(t) &= f_{x_0}(t^2)2t \approx \frac{1}{\sqrt{2\pi}} \frac{\left(\frac{\gamma_2\numfreqs^2}{2}-(\numfreqs-1)\right)t+\frac{\numfreqs-1}{2}}{\sqrt{2(\numfreqs-1)}\left(1-\frac{1}{t}\right)^{3/2}t^2} \notag \\ 
&\times \exp\left(-\frac{(\gamma_2\numfreqs^2t-(\numfreqs-1))^2}{4(\numfreqs-1)\left(1-\frac{1}{t}\right)}\right), \quad t > 0
\end{align}
\end{appendices}




\ifCLASSOPTIONcaptionsoff
  \newpage
\fi




\begin{thebibliography}{10}
\bibitem{liu22abb}
F.~Liu, Y.~Cui, C.~Masouros, J.~Xu, T.~X. Han, Y.~C. Eldar, and S.~Buzzi, ``Integrated sensing and communications: Toward dual-functional wireless networks for {6G} and beyond,'' \emph{IEEE J. Sel. Areas Commun.}, vol.~40, no.~6, pp. 1728--1767, 2022.

\bibitem{Wei23}
Z.~Wei, H.~Qu, Y.~Wang, X.~Yuan, H.~Wu, Y.~Du, K.~Han, N.~Zhang, and Z.~Feng, ``Integrated sensing and communication signals toward {5G-A} and {6G}: A survey,'' \emph{IEEE Internet Things J.}, vol.~10, no.~13, pp. 11\,068--11\,092, 2023.

\bibitem{kaushik2023isac}
A.~Kaushik, R.~Singh, S.~Dayarathna, R.~Senanayake, M.~Di~Renzo, M.~Dajer, H.~Ji, Y.~Kim, V.~Sciancalepore, A.~Zappone, and W.~Shin, ``Toward integrated sensing and communications for {6G}: Key enabling technologies, standardization, and challenges,'' \emph{IEEE Commun. Standards Mag.}, vol.~8, no.~2, pp. 52--59, 2024.

\bibitem{Han23local}
T.~Han, R.~Senanayake, P.~J. Smith, and J.~S. Evans, ``Local accuracy analysis of {FSK}-based joint radar and communications,'' in \emph{2023 IEEE Globecom Workshops (GC Wkshps)}, 2023, pp. 1147--1152.

\bibitem{Han25global}
T.~Han, P.~J. Smith, U.~Mitra, J.~S. Evans, and R.~Senanayake, ``Phase-optimised {FSK for ISAC},'' \emph{IEEE Trans. Wirel. Commun.}, 2025, to be published.

\bibitem{Levanon2004}
N.~Levanon and E.~Mozeson, \emph{Radar Signals}.\hskip 1em plus 0.5em minus 0.4em\relax John Wiley \& Sons, Inc, 2004.

\bibitem{Gaudio19}
L.~Gaudio, M.~Kobayashi, B.~Bissinger, and G.~Caire, ``Performance analysis of joint radar and communication using {OFDM} and {OTFS},'' in \emph{2019 IEEE Int. Conf. Commun. Workshops}, 2019, pp. 1--6.

\bibitem{Wang22}
Y.~Wang, Z.~Wei, W.~Zhou, K.~Han, and Z.~Feng, ``Triangular {FM-OFDM} waveform design for integrated sensing and communication,'' in \emph{2022 IEEE Int. Conf. Commun. Workshops (ICC Workshops)}, 2022, pp. 515--519.

\bibitem{Huang22}
Y.~Huang, S.~Hu, S.~Ma, Z.~Liu, and M.~Xiao, ``Designing low-{PAPR} waveform for {OFDM}-based {RadCom} systems,'' \emph{IEEE Trans. Wirel. Commun.}, vol.~21, no.~9, pp. 6979--6993, 2022.

\bibitem{Jiang08}
T.~Jiang and Y.~Wu, ``An overview: Peak-to-average power ratio reduction techniques for {OFDM} signals,'' \emph{IEEE Trans. Broadcast.}, vol.~54, no.~2, pp. 257--268, May 2008.

\bibitem{Piyush23}
P.~Varshney, P.~Babu, and P.~Stoica, ``Low-{PAPR} {OFDM} waveform design for radar and communication systems,'' \emph{IEEE Trans. Radar Syst.}, vol.~1, pp. 69--74, 2023.

\bibitem{richards14}
M.~Richards, \emph{Fundamentals of Radar Signal Processing}, 2nd~ed.\hskip 1em plus 0.5em minus 0.4em\relax McGrawHill Education, 2014.

\bibitem{Senanayake22TWCabb}
R.~Senanayake, P.~Smith, T.~Han, J.~Evans, W.~Moran, and R.~Evans, ``Frequency permutations for joint radar and communications,'' \emph{IEEE Trans. Wirel. Commun.}, vol.~21, no.~11, pp. 9025--9040, 2022.

\bibitem{Dayarathna23}
S.~Dayarathna, R.~Senanayake, P.~Smith, and J.~Evans, ``Frequency permutation subsets for joint radar and communication,'' \emph{IEEE Trans. Wirel. Commun.}, vol.~23, no.~2, pp. 1143--1157, 2024.

\bibitem{Han2023_freqpermPSK}
T.~Han, R.~Senanayake, P.~Smith, J.~Evans, W.~Moran, and R.~Evans, ``Combined radar and communications with phase-modulated frequency permutations,'' \emph{IEEE Open J. Commun. Soc.}, vol.~4, pp. 967--989, 2023.

\bibitem{Shi24}
M.~Shi, X.~Li, J.~Liu, and S.~Lv, ``Constant modulus waveform design for {RIS}-aided {ISAC} system,'' \emph{IEEE Trans. Veh. Technol.}, vol.~73, no.~6, pp. 8648--8659, 2024.

\bibitem{Bazzi23}
A.~Bazzi and M.~Chafii, ``On integrated sensing and communication waveforms with tunable {PAPR},'' \emph{IEEE Trans. Wireless Commun.}, vol.~22, no.~11, pp. 7345--7360, Nov. 2023.

\bibitem{greco18}
M.~S. Greco, F.~Gini, P.~Stinco, and K.~Bell, ``Cognitive radars: On the road to reality: Progress thus far and possibilities for the future,'' \emph{IEEE Sig. Process. Mag.}, vol.~35, no.~4, pp. 112--125, 2018.

\bibitem{Ghadian2020adaptive}
M.~Ghadian, R.~Fatemi~Mofrad, and B.~Abbasi~Arand, ``Fully adaptive waveform parameter design for cognitive tracking radars,'' \emph{IET Radar Sonar Navigation}, vol.~14, no.~10, pp. 1616--1623, 2020.

\bibitem{Steck18}
M.~Steck, C.~Neumann, and M.~Bockmair, ``Cognitive radar principles and application to interference reduction,'' in \emph{2018 19th Int. Radar Symp. (IRS)}, 2018, pp. 1--10.

\bibitem{Tsistrakis2014}
G.~Tsistrakis, ``Adaptive waveform design for cognitive radar,'' Ph.D. thesis, Heriot-Watt University, Edinburgh, Scotland, Dec. 2014, {Available}: \url{https://apps.dtic.mil/sti/citations/ADA284611}.

\bibitem{Ying2009}
Y.~Li, W.~Moran, S.~P. Sira, A.~Papandreou-Suppappola, and D.~Morrell, ``Adaptive waveform design in rapidly-varying radar scenes,'' in \emph{2009 Int. Waveform Diversity Des. Conf.}, 2009, pp. 263--267.

\bibitem{Zhang2018_adaptive}
X.~Zhang and X.~Liu, ``Adaptive waveform design for cognitive radar in multiple targets situation,'' \emph{Entropy}, vol.~20, no.~2, 2018.

\bibitem{Haykin2010}
S.~Haykin, A.~Zia, I.~Arasaratnam, and Y.~Xue, ``Cognitive tracking radar,'' in \emph{2010 IEEE Radar Conf.}, 2010, pp. 1467--1470.

\bibitem{prelcic2024survey}
N.~González-Prelcic, M.~F. Keskin, O.~Kaltiokallio, M.~Valkama, D.~Dardari, X.~Shen, Y.~Shen, M.~Bayraktar, and H.~Wymeersch, ``The integrated sensing and communication revolution for {6G}: Vision, techniques, and applications,'' \emph{Proc. IEEE}, 2024, early access.

\bibitem{palo2020}
F.~D. Palo, G.~Galati, G.~Pavan, C.~Wasserzier, and K.~Savci, ``Introduction to noise radar and its waveforms,'' \emph{Sensors}, vol.~20, no.~18, p. 5187, 2020.

\bibitem{akan2020}
O.~B. Akan and M.~Arik, ``Internet of radars: Sensing versus sending with joint radar-communications,'' \emph{IEEE Commun. Mag.}, vol.~58, no.~9, pp. 13--19, 2020.

\bibitem{Ali24}
I.~Ali, H.~Zahan, and S.~Mastorakis, ``Sensor clouds: Recent advancements, use cases and open challenges,'' \emph{IEEE Internet Things Mag.}, vol.~7, no.~1, pp. 98--103, 2024.

\bibitem{Wu23lpwan}
D.~Wu, A.~S. Bogdan, and J.~Liebeherr, ``Large-scale environmental sensing of remote areas on a budget,'' \emph{IEEE Internet Things Mag.}, vol.~6, no.~2, pp. 130--136, 2023.

\bibitem{de20195g}
I.~B.~F. De~Almeida, L.~L. Mendes, J.~J. Rodrigues, and M.~A. Da~Cruz, ``{5G} waveforms for {IoT} applications,'' \emph{IEEE Commun. Surv. Tut.}, vol.~21, no.~3, pp. 2554--2567, 2019.

\bibitem{Paulose1994}
A.~Paulose, ``High radar range resolution with the step frequency waveform,'' M.S. thesis, Naval Postgraduate School, Monterey, CA, Jun. 1994, {Available}: \url{https://apps.dtic.mil/sti/citations/ADA284611}.

\bibitem{Soares1996}
P.~A. Soares, ``Stepped frequency waveform design and analysis using the ambiguity function,'' M.S. thesis, Naval Postgraduate School, Monterey, CA, Jun. 1996, {Available}: \url{https://apps.dtic.mil/sti/citations/ADA315096}.

\bibitem{Kozlov22}
R.~Kozlov, K.~Gavrilov, T.~Shevgunov, and V.~Kirdyashkin, ``Stepped-frequency continuous-wave signal processing method for human detection using radars for sensing rooms through the wall,'' \emph{Inventions}, vol.~7, no.~3, 2022.

\bibitem{Proakis08}
J.~G. Proakis and M.~Salehi, \emph{Digital Communications}.\hskip 1em plus 0.5em minus 0.4em\relax NY, USA: McGraw-Hill,, 2008.

\bibitem{vantrees01}
H.~L. VanTrees, \emph{Detection, {E}stimation, and {M}odulation {T}heory: {R}adar-{S}onar {P}rocessing and {G}aussian {S}ignals in {N}oise}.\hskip 1em plus 0.5em minus 0.4em\relax NY, USA: John Wiley \& Sons, Inc, 2001.

\bibitem{Lawal1980}
H.~B. Lawal and G.~J.~G. Upton, ``An approximation to the distribution of the $\chi^2$ goodness-of-fit statistic for use with small expectations,'' \emph{Biometrika}, vol.~67, no.~2, p. 447–453, 1980.

\bibitem{Ferebee82}
B.~Ferebee, ``The tangent approximation to one-sided {B}rownian exit densities,'' \emph{Z. Wahrscheinlichkeitstheorie verw. Gebiete 61}, vol.~61, pp. 309--326, 1982.

\bibitem{Tsonev11}
D.~Tsonev, S.~Sinanovic, and H.~Haas, ``Enhanced subcarrier index modulation {(SIM) OFDM},'' in \emph{2011 IEEE GLOBECOM Workshops (GC Wkshps)}, 2011, pp. 728--732.

\end{thebibliography}
%



\end{document}